\documentclass[12pt]{article}
\pdfoutput=1

\usepackage{float}

\usepackage[table]{xcolor}

 \usepackage[a4paper,
   left=2.5cm, right=2.5cm,
   top= 3cm, bottom=4cm]{geometry}
\usepackage{amsmath,amssymb,amsfonts}
\usepackage{hyperref}
\usepackage{bbm} 
\usepackage{bm} 
\usepackage[toc,page]{appendix}
\usepackage{cite}
\usepackage{url}

\numberwithin{equation}{section}

\usepackage{makecell,multirow}
\usepackage{pgfplots}
\pgfplotsset{compat = newest}

\newcommand{\rem}[1]{}

\def\Z{\mathbb{Z}}
\def\Q{\mathbb{Q}}
\def\R{\mathbb{R}}

\def\P{\mathbb{P}}

\def\rk{\operatorname{rk}}

\def\Hirz[#1]{\mathbbm{F}_{#1}}
\def\o[#1]{\overline{#1}}

\def\Nf{N_{\rm flux}}

\def\Ktt{\mbox{K3} \times \mbox{K3}}
\def\K3{\mbox{K3}}

\DeclareMathSymbol{\shortminus}{\mathbin}{AMSa}{"39}

\newcommand{\tors}{\mbox{tors}}

\def\Re{\mbox{Re}}
\def\Im{\mbox{Im}}

\newcommand{\beq}{\begin{equation}}
\newcommand{\eeq}{\end{equation}}

\newtheorem{THM}{Theorem}[section]

\newtheorem{PR}{Proposition}[section]

\begin{document}

\begin{titlepage}

\vspace*{-2cm} 
\begin{flushright}

{\tt preprint} \qquad \qquad 

\end{flushright}

\vspace*{0.8cm} 
\begin{center}
{\Huge Tadpoles and Gauge Symmetries}\\

 \vspace*{1.5cm}
Andreas P. Braun,$^1$ Bernardo Fraiman,$^{2}$ Mariana Gra\~na,$^3$ Severin L\"ust$^4$\\ and H\'ector Parra De Freitas$^3$\\

 \vspace*{1.0cm} 
$^1$ {\it Department of Mathematical and Computing Sciences,  Durham University Upper Mountjoy Campus, Stockton Rd, Durham DH1 3LE, UK}\\[2mm]

$^2$ {\it CERN, Theoretical Physics Department, 1211 Meyrin, Switzerland}\\[2mm]

$^3$ {\it Institut de Physique Th\'eorique, Universit\'e Paris Saclay, CEA, CNRS Orme des Merisiers, 91191 Gif-sur-Yvette CEDEX, France}\\[2mm]

$^4$ {\it Laboratoire Charles Coulomb (L2C), Université de Montpellier, CNRS, F-34095, Montpellier, France}\\[4mm]
{\tt {andreas.braun@durham.ac.uk, bernardo.fraiman@cern.ch, mariana.grana@ipht.fr, severin.lust@umontpellier.fr, hector.parradefreitas@ipht.fr}} \\

\vspace{1cm}
\small{\bf Abstract} \\[3mm]\end{center}
The tadpole conjecture proposes that complex structure moduli stabilisation by fluxes that have low tadpole charge can be realised only at special points in moduli space, leading generically to (large) gauge symmetries. Here we provide an exhaustive survey of the gauge symmetries arising in F-theory flux compactifications on products of attractive $\K3$ surfaces, with complex structure moduli fully stabilised. We compute the minimal rank of the left-over non-abelian gauge group for all flux configurations within the tadpole bound, finding that it is always non-zero. It decreases in a roughly linear fashion with the tadpole charge, reaching zero at charge 30. By working out possible gauge algebras for different values of the tadpole, we find that all simple ADE Lie algebras of rank $\le 18$ appear.

\end{titlepage}

\newpage

\tableofcontents

\section{Introduction}
Flux compactifications in type IIB orientifolds and, more generally, F-theory, make up the bulk of the string landscape explored in the last two decades. A tacit assumption that underlies 
much of the work on the subject is that it is possible to make a sufficiently `generic' choice of flux which stabilises all complex structure moduli. Recent 
work examining complex structure moduli stabilisation together with the interplay of flux quantisation and the tadpole constraint has begun to challenge this point of view. The basic picture is that flux quantisation says that allowed fluxes sit in a lattice, whereas the tadpole constraint limits the norm of lattice vectors that can be chosen, so that sufficiently generic choices of \emph{quantised} fluxes might surpass the maximum permitted by the tadpole.
Generic choices of fluxes however do not preserve any supersymmetry, which requires a self-duality property. On the other hand, there typically are simple choices flux that satisfy the 
supersymmetry conditions, and stay within the tadpole bound. Nonetheless, these generically either leave a moduli space of possible complex structures, or if they completely fix complex structure moduli, they do so at special (symmetric) points in moduli space (see e.g. Refs \cite{Giryavets:2003vd,Cicoli:2013cha,Demirtas:2019sip,Braun:2020jrx,Lust:2022mhk}  for stabilisation at symmetric points).\footnote{Note however that the constructions in type IIB orientifold compactifications require also the negative D3 charge coming from D7 branes wrapped on four-cycles (see \cite{Crino:2022zjk} for an exhaustive analysis of tadpole charge coming from O3 planes, as well as D7 branes and O7 planes in Calabi-Yau three-folds in the  Kreuzer-Skarke list). These however come with a large number of  moduli of their own, whose stabilisation is not taken into account. It is very hard, if not impossible, to stabilise all D7-moduli  within the bound \cite{Bena:2021qty}. In the F-theory picture complex structure  and D7-moduli are unified into complex structure moduli of the four-fold.}

As far  as the stabilisation  of (a large number of) complex structure moduli at {\it generic points} is concerned, Ref. \cite{Bena:2020xrh} proposed that there is a universal bound limiting the number of stabilised moduli as a function of the tadpole charge. The precise ``tadpole conjecture"   is that the charge induced by fluxes that stabilise a large number of moduli at a generic point in moduli space is larger than 1/3 of the number of stabilised moduli. On the other hand, the tadpole bound limits this charge to 1/4 times the total number of complex structure moduli, such that one cannot stabilise all moduli within the bound.  
Besides existing examples in the literature, the conjecture was supported by an exploration in the $\Ktt$ lattice (as well as in smaller-dimensional ones) using evolutionary algorithms \cite{Bena:2021wyr}, and later given strong evidence in the large complex structure limit \cite{Plauschinn:2021hkp,Tsagkaris:2022apo,Grana:2022dfw,Coudarchet:2023mmm} and also in non-geometric compactifications \cite{Becker:2022hse}. On the other hand, the conjecture  has been challenged by the ``linear scenario" mechanism of moduli stabilisation \cite{Marchesano:2021gyv}; however, there is a potential loophole in this analysis \cite{Lust:2021xds}.

The evolutionary algorithms used in \cite{Bena:2020xrh, Bena:2021wyr} provided ${\cal O}(10^5)$ different choices of fluxes stabilising moduli at a generic point in the moduli space of $\Ktt$ compactifications with a charge induced that exceeds by one the tadpole bound of 24 (in units of M2-brane charge), while no example within the bound was found. This strongly suggests that  one can stabilise moduli only at special points within the bound. For the case at hand, these points are such that there is a left-over non-Abelian gauge symmetry, carrying with it extra massless scalar fields. This raises the very interesting question that we investigate in this paper, namely what is the interplay between the appearance of non-Abelian gauge symmetries at specific points in moduli space and the induced tadpole charge of the fluxes stabilising moduli at that specific point.

In this paper we answer this question for $\Ktt$ compactifications of F-theory. Our results show that in vacua with all complex structure moduli perturbatively stabilised in a supersymmetric minimum, the tadpole bound enforces non-trivial gauge theory sectors. Our analysis has two crucial ingredients. Flux solutions for 
M-theory on $\Ktt$ where all complex structure moduli are stabilised can be found using the approach of Aspinwall and Kallosh \cite{Aspinwall:2005ad}. This 
results in both $\K3$ surfaces being attractive, i.e. they both have a Picard lattice of the maximal rank, $20$. The enumeration of solutions becomes a problem in arithmetic, and all solutions within the tadpole bound were listed in \cite{Braun:2014ola}. 

These M-theory flux vacua have an F-theory description upon specifying an elliptic fibration on one of the two $\K3$ surfaces. The F-theory gauge group can then be read off from the frame lattice of the elliptic fibration, which encodes singular fibres and sections. For each flux solution, the frame lattices  of all elliptic fibrations can be determined by studying embeddings of an auxiliary lattice into the 24-dimensional Niemeier lattices. 

We design a computer algorithm that explores all possible embeddings, and determines the rank of the resulting gauge group in F-theory. We explore all solutions up to the tadpole bound, and find that there is always a non-Abelian gauge group. Furthermore, the minimum charge where moduli can be stabilised at a point where there is no left-over non-Abelian gauge group (i.e., a point that is referred to in \cite{Bena:2020xrh} as generic) is $30$. We list the minimum rank of the gauge groups for most of the solutions (some Picard lattices are much harder to explore, and for those we only answer the yes-no question of whether there can be no non-Abelian gauge groups). We also give some of the gauge groups for the simpler Picard lattices.

The paper is organised as follows: in Section \ref{sec:fluxvacua} we review flux vacua on $\Ktt$, presenting the M-theory solutions of \cite{Aspinwall:2005ad}, and the method to find their elliptic fibrations, while in section \ref{sec:results} we present our results. In Appendix \ref{app:lattices} we give an extensive review of lattices, introducing all the concepts needed in the paper.

\section{M- and F-theory flux vacua on \texorpdfstring{$\Ktt$}{K3 x K3}}
\label{sec:fluxvacua}

In this section we review flux vacua of M- and F-theory on $X = S_1 \times S_2$, for $S_1$ and $S_2$ 
$\K3$ surfaces. Technical details of this material can be found in \cite{Dasgupta:1999ss,Tripathy:2002qw,Moore:1998pn,Aspinwall:2005ad,Braun:2008pz,
Braun:2013yya,Braun:2014ola}, see also \cite{Denef:2008wq} for a general review on F-theory flux vacua.

\subsection{M-theory flux vacua on Calabi-Yau fourfolds}

M-theory on Calabi-Yau fourfolds $X$ allows the introduction of $G_4$ fluxes, while leaving the metric Calabi-Yau up to a conformal factor \cite{Becker:1996gj}, and which 
generate a superpotential that depends on the location in complex structure moduli space \cite{Gukov:1999ya}. The $G_4$ fluxes obey a quantisation condition
\beq \label{G4quant}
G_4  + \frac{c_2(X)}{2} \in H^4(X,{\mathbb Z})\ ,
\eeq
and are subject to the tadpole constraint
\beq \label{tadpole}
N_{\rm flux}  + N_{M2} =  \frac{\chi(X)}{24} \, , 
\eeq
where $N_{M2}$ is the number of M2 branes in the space-time transverse to $X$ and
\beq \label{Nf}
\Nf=\frac{1}{2} \int_X G_4 \wedge G_4
\eeq
is the M2-brane charge of the fluxes.

For supersymmetric Minkowski minima the complex structure moduli of the four-fold are such that \cite{Becker:1996gj} 
\beq \label{susyflux}
G_4 \in H^{2,2}_{\rm{prim}} (X) \,,
\eeq
in other words fluxes must be of Hodge type $(2,2)$ and primitive: $G_4 \wedge J = 0$, where $J$ is the K\"ahler form on $X$. Depending on how much fluxes are turned on, the $(2,2)$ requirement can fix some or all of the complex structure moduli of the Calabi-Yau 4-fold. The primitivity condition additionally constrains the K\"ahler moduli. 
Furthermore, $N_{M2}$ must be positive for supersymmetric solutions. As $G_4 \in H^{2,2}_{\rm{prim}} (X)$ is self-dual, it turns $\int_X G_4 \wedge G_4$ into a positive number, and thus  there can only be finitely many flux choices for a fixed point in moduli space which are bounded by the Euler characteristic of the fourfold $X$.

\subsection{F-theory flux vacua on Calabi-Yau fourfolds}

When the four-fold $X$ carries an elliptic fibration 
\begin{equation}
\begin{aligned}
E \hookrightarrow &X& \\ 
 &\downarrow&\\
 &B&
\end{aligned}
\end{equation}
one can take the limit in which the volume of $E$ goes to zero. Working in F-theory, one may assume without loss of generality that the fibration is described by 
(a resolution of) a Weierstrass model.\footnote{Technically, this can be accomplished by passing to the associated Jacobian fibration and resolving singularities.} The limit of vanishing fibre is dual to the compactification of F-theory on $X$ (employing the elliptic fibration by $E$) and yields a Lorentz invariant theory in four dimensions if the flux has `one leg on the fibre', a condition that is equivalent to demanding that the integral of $G_4$ over certain divisors of $X$ 
vanishes \cite{Grimm:2011fx}.  

The quantisation, tadpole and supersymmetry conditions for F-theory fluxes are the same as in M-theory (Eqs \eqref{G4quant}-\eqref{susyflux}).  The difference in the latter is that F-theory fluxes, having one leg on the fiber, are automatically primitive in manifolds of strict SU(4) holonomy, and therefore do not stabilise K\"ahler moduli. In this setup K\"ahler moduli are only stabilised by non-perturbative corrections \cite{Kachru:2003aw}, which we do not consider here. 

Regarding complex structure moduli stabilisation in F-theory flux compactifications, it was conjectured in \cite{Bena:2020xrh} that stabilisation of a large number of  moduli at a {\it generic point} in moduli space requires a flux charge $N_{\rm flux}$ which grows linearly with the number of moduli. Furthermore, the coefficient of the linear growth was conjectured to be larger than $\tfrac13$, while $\tfrac{\chi}{24} \sim \tfrac14 h^{3,1}$ and thus a large number of complex structure moduli cannot be stabilised at a generic point in moduli space within the tadpole bound. If the tadpole conjecture is true, then stabilising moduli within the tadpole will force upon special points or regions in moduli space which have some sort symmetry, or there are singularities. In F-theory, singularities of $X$ lead to non-Abelian gauge groups, massless matter representations, and Yukawa couplings, depending on their codimension in $X$. 

The tadpole conjecture hence implies the existence of non-trivial gauge theory sectors in models with completely stabilised complex structure moduli. This is an intriguing possibility, as it goes against the naive expectation that flux vacua populate generic points in moduli space, at which no non-abelian gauge groups (beyond non-Higgsable clusters \cite{Morrison:1996pp,Morrison:2012np}) reside.\footnote{See \cite{Braun:2014xka,Braun:2014lwp} for estimates of the statistical cost of non-abelian gauge groups.}

In what follows, we show that this is indeed the case for $\Ktt$ compactifications with a particularly simple flux. Since $\Ktt$ does not have strict SU(4) holonomy, fluxes can in principle stabilise all moduli, as explored in \cite{Bena:2020xrh}. Here, we will however restrict to a subset of all possible fluxes  that are amenable to an exhaustive analysis, which can only stabilise complex structure moduli.

\subsection{M- and F-theory on \texorpdfstring{$\Ktt$}{K3 x K3}}
\label{sec:MFK3}

Analysing what is the shortest integral flux that stabilises all complex structure moduli at a given point in the moduli space of a Calabi-Yau fourfold, let alone throughout, is a daunting task and most work has focussed on the large complex structure limit \cite{Plauschinn:2021hkp,Tsagkaris:2022apo,Grana:2022dfw}.  Here, we  focus on the specific case of $X = S_1 \times S_2$ for a pair of $\K3$ surfaces $S_1$ and $S_2$, where we can be significantly more precise. 

A $\K3$ surface is the unique non-trivial Calabi-Yau manifold in complex dimension two (see \cite{barth2012compact,Aspinwall:1996mn,huybrechts_2016} for an in-depth discussion). 
Any element $\delta_C$ of the integral middle cohomology of a $\K3$ surface $S$ is dual to a curve $C$, and the inner form is related to the Euler characteristic of $C$ by
\begin{equation}
-\chi(C) = \int_S  \delta_C \wedge \delta_C \equiv \delta_C \cdot \delta_C\,, 
\end{equation}
which implies that $\delta_C \cdot \delta_C$ is an even integer. Picking a $\Z$-basis of the middle cohomology composed of elements $\delta_I$, $I=1,...,22$, Poincar\'e duality then implies that the inner form 
\begin{equation}
d_{IJ} = \delta_I \cdot \delta_J 
\end{equation}
defines an even unimodular lattice. By the Hirzebruch signature theorem the signature of this lattice is $(n,n+16)$, which together with $b^2(S) = 22$ implies that $n=3$. Such lattices are covered by strong classification theorems (see Appendix \ref{app:lattices} for an extensive review of lattices) and one finds that they are all isomorphic to
 \beq \label{dK3}
 \Lambda_{3,19} = U \oplus U \oplus U \oplus E_8[-1] \oplus E_8[-1] \ , 
 \eeq
where $E_8[-1]$ is minus the Cartan matrix of $E_8$ and $U$ is the $2\times2$ matrix defined in \eqref{U}.

Not only do we know rather explicitly what the integral middle cohomology of a $\K3$ surface is like, but we have a complete picture of the moduli space due to the global Torelli theorem. It says that the moduli space of Ricci-flat metrics on a $\K3$ (which is equal to the moduli space of M-theory on $\K3$) is given by the coset 
\begin{equation}
O(\Lambda^{3,19}) \backslash O(3,19)/O(3) \times O(19) ~ \times {\mathbb R}^+ 
\end{equation}
that parameterises the deformations of a 3-plane $\Sigma$ of positive-norm vectors inside a 22-dimensional vector space modulo automorphisms $O(\Lambda^{3,19})$ 
of $\Lambda^{3,19}$, together with the volume of the $\K3$. 

The three vectors $\omega_a \in H^2(S,{\mathbb R})$, $a=1,2,3$, spanning the three-plane $\Sigma$ define a hyper-K\"ahler structure, and for M-theory compactifications on $S_1 \times S_2$, hyper-K\"ahler rotations give rise to the R symmetries of the resulting 3D $\mathcal{N} = 4$ theory. 
Choosing a complex structure, we can write the K\"ahler form and holomorphic 2-form as
\beq
\Omega= \omega_1 + i \omega_2 \ , \quad J=\sqrt{2 {\rm vol}(S)} \, \omega_3\, .
\eeq
In this complex structure, the space $H^{1,1}(S,{\mathbb C})$ is the 20-dimensional space orthogonal to $\omega_1$ and $\omega_2$. The intersection of this hyper-plane and the lattice of integral 2-forms defines a sublattice whose rank can be at most 20, called the Picard lattice
 \begin{equation} \label{Pic}
Pic(S) := H^{1,1}(S) \cap H^2({S,\mathbb{Z}})  \, .
 \end{equation}
which is (dual to) the lattice of holomorphic curves. This lattice has always signature $(1,r)$. 

There are special points (at finite distance) in the moduli space where the $\K3$ surface develops ADE singularities. For a curve $C$ isomorphic to a $\P^1$ 
(so that $C^2 =-2$), there always exists a representative with minimal volume in its homology class given by
\begin{equation}
\mbox{Vol}^2(C) = \sum_a \left(\int_C \omega_a\right)^2 \, ,
\end{equation}
so that curves perpendicular to $\Sigma$ have collapsed to zero volume. This signals the appearance of ADE singularities. For a given choice of $\Sigma$, we can characterize the singularities by studying the lattice generated by all vectors with norm $-2$ contained in $\Sigma^\perp$:
\begin{equation}\label{lambda_s}
\Lambda_S = (\Sigma^\perp)_\text{root} \equiv\langle v \in H^2(S,\Z) | v^2 = -2, v \cdot \Sigma = 0 \rangle  \, .
\end{equation}
Hence $\Lambda_S[-1]$ is a root lattice generated by elements which square to $2$, so that we can write
\begin{equation}
\Lambda_S[-1] = \Gamma_1 \oplus \Gamma_2 \oplus \cdots
\end{equation}
for $\Gamma_k$ ADE root lattices. The inner form on each $\Gamma_k$ is the Cartan matrix of an associated Lie algebra $\mathfrak{g}_k$, and for each ADE summand, the associated $\K3$ surface carries one instance of the corresponding ADE singularity. For M-theory on a $\K3$ surface, the non-abelian gauge algebra is hence 
\begin{equation}
\mathfrak{g} = \mathfrak{g}_1 \oplus \mathfrak{g}_2 \oplus \cdots \ .
\end{equation}

M-theory compactifications on $S_1 \times S_2$ have an F-theory uplift if one of the $\K3$ (say, $S_1$) has an elliptic fibration. Reducing M-theory on one of the directions of the elliptic fibration, sending its volume  to zero and taking the T-dual, one gets a type IIB compactification to four dimensions on $S_2$ times the ${\mathbb P}^1$ base of the elliptic fibration, with a varying axion-dilaton. The latter is the complex structure of the elliptic fibration. This is better described in terms of F-theory, again on $S_1 \times S_2$, where $S_1$ is elliptically fibered with a fixed volume.  On an elliptically fibered $\K3$ there are at least two algebraic curves, one corresponding to the $T^2$ fiber, and the other to the section, which is equivalent to the base ${\mathbb P}^1$. Given these topologies and the fact that a section meets every fibre exactly once implies that the homology classes of base and the fiber span a lattice with inner form 
\begin{equation}
\begin{pmatrix}
-2 & 1 \\
 1 & 0
\end{pmatrix}
\end{equation}
which is equivalent to $U$ modulo $SL(2,\Z)$. Thus, for every elliptic fibration there is a copy of $U$ embedded in the Picard lattice of the $\K3$ surface. The converse of this statement is true as well: for any primitive embedding $U \hookrightarrow Pic(S)$ there exists an associated elliptic fibration with a section. The unimodularity of $U$ allows us to write
\begin{equation}
Pic(S) = U \oplus W \ ,
\end{equation}
where $W$ is called the frame lattice of the elliptic fibration. 

The frame lattice $W$ contains a great deal of information on the singular fibres and further sections of the elliptic fibration. Working with a smooth model,\footnote{In M-theory, we can resolve any elliptic fibration keeping the complex structure fixed.} an elliptic fibration on a $\K3$ surface can have a collection of reducible fibres over points in the $\P^1$ base, each of which can be decomposed into a collection of $\P^1$s arranged according to an extended ADE Dynkin diagram. The affine node is distinghuished as the component of the fibre that is met by the section of the elliptic fibration. The root sublattice $W_\text{root}$ of $W$ is generated by all of the remaining fibre components. Passing to a Weierstrass model, all these fibre components are collapsed, so that both $\Omega$ and the K\"ahler form $J$ are perpendicular to $W$ ($J$ only takes values in $U \otimes \R$ in this case). We can read off the ADE singularities of the Weierstrass model and hence the F-theory gauge algebra from (c.f. eq. \eqref{lambda_s})
\begin{equation}
\Lambda_S = W_\text{root} \equiv \langle v \in W | v^2 = -2 \rangle  \, .
\end{equation}

Generators of $W$ which are not roots correspond to extra sections of the elliptic fibration and we can write 
\begin{equation}
MW(S) = W/W_\text{root} 
\end{equation}
where $MW(S)$ is the Mordell-Weil group of $S$. Note that $MW(S)$ is in general not a lattice but can contain torsional elements. These correspond geometrically 
to torsional sections and determine the global form of the F-theory gauge group \cite{Aspinwall:1998xj,Mayrhofer:2014opa}. 

In a type IIB picture, the occurence of non-Abelian gauge algebras of ADE type is due to stacks of $(p,q)$ 7-branes located at the points of the base $\P^1$ of the elliptic $\K3$ surface over which the fibre degenerates. For F-theory on $S_1 \times S_2$, with $S_1$ elliptically fibered, 
the $(p,q)$ 7-branes wrap $S_2$ entirely, but do not have any mutual intersections. This makes it clear that F-theory on $S_1 \times S_2$ has no bifundamental 
charged matter, but only charged fields transforming in the adjoint.

\subsection{Flux vacua on \texorpdfstring{$\Ktt$}{K3 x K3}}
\label{sec:MFfluxK3}

Let us now describe flux vacua on $S_1 \times S_2$ in M-theory. As we are ultimately interested in flux backgrounds that have an F-theory lift, $G_4$ cannot have components proportional to the volume forms of $S_1$ or $S_2$, so that $G_4$ is a sum of wedge products of integral 2-forms on each $\K3$. Forms of Hodge type $(2,2)$ are obtained either by wedging $(1,1)$ forms on each $\K3$, or the $(2,0)$ with the $(0,2)$ form of each of the $\K3$, namely
\begin{equation} \label{genG4}
G_4 = \sum_{I,J} c_{IJ }\eta_I \wedge \eta_J  + {\rm Re} \left( \gamma \Omega_1 \wedge \bar \Omega_2 \right)\,,
 \end{equation}
where $\eta_I, \eta_J$ are $(1,1)$-forms on each $\K3$, $\Omega_1$ and $\Omega_2$ are the (2,0) forms and the $c_{IJ}$ are chosen such that $G_4$ is primitive. For an F-theory lift, we need to make sure that at least one of the $\K3$ surfaces has an elliptic fibration and that $G_4$ has vanishing wedge product with the fibre and base of the elliptic fibration. While these conditions are automatically satisfied for the last component of the flux, they have to be imposed by hand on the other components. The quantisation condition is that $G_4 \in H^4(S_1 \times S_2, \Z)$, as $c_2(S)$ is even for any $\K3$ surface. As the integral homology of a
$\K3$ surface has no torsion, it follows that $H^4(S_1 \times S_2, \Z) = H^2(S_1,\Z) \otimes H^2(S_2,\Z)$ by using the integral version of the K\"unneth theorem \cite{dodson_parker_97}. 

Note that writting $G_4$ as \eqref{genG4}, only the sum of the two pieces needs to be integral, but they can in principle be non-integral individually. 
Finally, we need to make sure that the tadpole constraint 
\begin{equation}
\Nf=\frac12 \int_{S_1 \times S_2} G_4 \wedge G_4 \leq 24 \, 
\end{equation}
is satisfied. 

As shown in \cite{Aspinwall:2005ad} choices for which $G_4$ is purely of the type
\begin{equation}\label{eq:fluxpuretype0}
G_4 = {\rm Re} \left( \gamma \Omega_1 \wedge \bar \Omega_2 \right) \,,
\end{equation}
with $\gamma$ a constant appropiately chosen to ensure quantisation (see below), give rise to 3D $\mathcal{N} =2$ Minkowski vacua where all deformations of $\Omega_1$ and $\Omega_2$ are fixed, while leaving $J$ perturbatively unfixed.\footnote{Non-perturbative corrections generically generate a superpotential for these \cite{Aspinwall:2005ad}.} We will restrict to such solutions from now on, mainly for two reasons. Most importantly, this is the closest analogue of an F-theory compactification on a manifold of strict SU(4) holonomy, where fluxes (perturbatively) only fix complex structure moduli. Furthermore, the existence of an elliptic fibration and the subsequent F-theory limit require us to take a particular limit of the K\"ahler form, which in turn limits the possible choices of $c_{IJ}$ in \eqref{genG4}. 

Note that for $\Ktt$ compactifications, the hyper-K\"ahler rotations of the two $\K3$ surfaces (corresponding to the R symmetries of the 3D $\mathcal{N}=4$ theory from M-theory compactification) prevent us from discriminating between complex structure and K\"ahler deformations for a given Ricci-flat metric. For a given metric, we can however always fix a specific complex structure and subsequently write the most general flux which results in supersymmetric Minkowski vacua as 
\eqref{genG4}. For the reasons explained above, we will further limit ourselves to fluxes of the form \eqref{eq:fluxpuretype0}. For every such choice the only condition for these M-theory flux solutions to have an F-theory dual is that one of the $\K3$ surfaces is elliptically fibered. As we will see in the follwing, this is always the case, i.e. every flux of this type leads to several F-theory vacua which are obtained by specifying an elliptic fibration.

We now describe the flux solutions \eqref{eq:fluxpuretype0} in some more detail (for further details and proof of the statements, see \cite{Aspinwall:2005ad}). Integrality of the flux requires the complex number $\gamma$ to enforce 
\begin{equation}\label{eq:fluxquantomega}
\Re \left( \gamma \Omega_1 \wedge \bar \Omega_2 \right) \in H^4(X,\mathbb{Z})\, .
\end{equation} 
Such a $\gamma$ and hence a corresponding M-theory solution exists under the following two conditions:
\begin{enumerate}
\item Both $\K3$ surfaces need to be {\bf attractive}, i.e. they are both at a point in their moduli spaces where the Picard lattice, defined in \eqref{Pic}, has rank 20. The
rank of the transcendental lattices 
\beq
T_{S_i} := Pic^\perp \subset \Lambda_{3,19} 
\eeq
is hence two. Denoting the generators or $T_{S_i}$ by $p_i,q_i$ we can write 
its inner form as
\begin{equation} \label{abc}
\begin{pmatrix}
p_i \cdot p_i & p_i \cdot q_i \\
p_i \cdot q_i  & q_i \cdot q_i
\end{pmatrix}
=
\begin{pmatrix}
2a_i & b_i \\
b_i & 2c_i 
\end{pmatrix} 
\simeq [a_i,b_i,c_i]\,,
\end{equation}
where we introduced the abbreviation $[a,b,c]$. By explicitely constructing the associated $\K3$ surfaces, \cite{shioda_inose_1977} showed that the converse to this is true as well, i.e. there exists an associated attractive $\K3$ surface for every such positive definite $T_{S}$. Furthermore this $\K3$ surface, i.e. the embedding of 
$T_S \hookrightarrow \Lambda_{3,19}$ is unique up to isometry. As we will discuss below, every attactive $\K3$ surface admits an elliptic fibration and hence an F-theory limit, so that all of the solutions discussed here lift to F-Theory.

\item Denoting the determinant of the inner form on $T_{S_i}$ by $Q_i := 4 a_i c_i- b_i^2$, the product $Q_1 Q_2$ must be a perfect square:
\begin{equation}\label{eq:square_cond}
Q_1 Q_2 = k^2 \hspace{1cm} \mbox{for} \hspace{1cm} k \in \Z. 
\end{equation}
\end{enumerate}

We can now work out the details of these solutions. As the $\K3$ surfaces are both attractive, one can write 
\begin{equation}
\Omega_i = p_i + \tau_i q_i 
\end{equation}
with 
\begin{equation}
 \tau_i = \frac{-b_i + i \sqrt{Q_i}}{2c_i}\,, 
\end{equation}
which makes the freezing of complex structure moduli manifest. 

The condition $Q_1 Q_2 = k^2$ for $k \in \Z$ implies that the two $\tau_i$ live in the same field extension of $\mathbb{Q}$, which implies that 
the complex number $\gamma$ can simultaneously satisfy the conditions
\begin{equation}\label{eq:gamma_conditions}
\begin{aligned}
\Re (\gamma) \in \Z\,, &\hspace{1cm}& \Re (\gamma \tau_1 ) \in \Z\,, \\
\Re (\gamma \bar{\tau}_2) \in \Z\,, && \Re (\gamma \tau_1 \bar{\tau}_2) \in \Z\,. \\
\end{aligned}\, 
\end{equation}
The induced tadpole can then be computed as
\begin{equation}
\Nf = \frac{1}{2} \int_X G_4 \wedge G_4 =  \frac{|\gamma|^2 k^2}{4 c_1 c_2}\, .
\end{equation}

\subsection{Examples}

Before explaining how to find all solutions up to a given tadpole, let us construct a number of solutions by hand. 
This is done by making specific choices for the integers $[a_1,b_1,c_1]$ and $[a_2,b_2,c_2]$ specifying a pair of attractive $\K3$ surfaces 
such that \eqref{eq:square_cond} is fulfilled.

\subsubsection{\texorpdfstring{$[a, 0, a] - [b, 0, b]$}{[a, 0, a] - [b, 0, b]}}

Choosing $T_{S_1} = [a, 0, a]$ is compatible with any $T_{S_2} = [b, 0, b]$ as
\begin{equation}
Q_1 Q_2 = 4a^2 \,\, 4 b^2 = (4ab)^2 \, .
\end{equation}
In this case
\begin{equation}
\tau_1 = \tau_2 = i 
\end{equation}
and a choice for $\gamma$ with minimal length that obeys \eqref{eq:gamma_conditions} is 
\begin{equation}
\gamma = 1\,. 
\end{equation}
The induced tadpole is then
\begin{equation}
\Nf = 4 a b \, .
\end{equation}

\subsubsection{\texorpdfstring{$[a, a, a] - [b, b, b]$}{[a, a, a] - [b, b, b]}}

Choosing $T_{S_1} = [a, a, a]$ is compatible with $T_{S_2} = [b, b, b]$ as
\begin{equation}
Q_1 Q_2 = 3a^2 \,\, 3 b^2 = (3ab)^2 \, .
\end{equation}
In this case
\begin{equation}
\tau_1 = \tau_2 = \frac{1}{2}(-1 +i \sqrt{3}) 
\end{equation}
and a choice for $\gamma$ with minimal length that obeys \eqref{eq:gamma_conditions} is 
\begin{equation}
\gamma = \frac{2i}{\sqrt{3}} \,.
\end{equation}
The induced tadpole is then
\begin{equation}
\Nf = 3 a b \, .
\end{equation}

\subsection{All M-theory solutions}

All solutions with $\Nf\le 24$ were obtained in \cite{Aspinwall:2005ad,Braun:2014ola}. Here we explain the method used and 
extend the results up to $\Nf=30$.
 
By exploiting the $SL(2,\Z)$ action on $T_{S_i}$, we can bound the values of $a_i,b_i,c_i$ such that
\begin{equation}
|b_i| \leq |c_i| \leq |a_i| \, .
\end{equation}
Furthermore, the positive definiteness of the inner form on $T_{S_i}$ implies that $a_i$ and $c_i$ are both positive. 
Finally, we can also assume that $b_i$ is non-negative. If $b_i$ is negative, we can map to positive $b_i$ 
by letting $q_i \rightarrow -q_i$ so that the same lattice is generated. 

To show that only finitely many solutions exist, we can use the tadpole to bound possible $a_i$. Let us first assume 
that $\Re(\gamma) \neq 0$. Then \eqref{eq:fluxquantomega} implies $\Re(\gamma) \in \Z $ which means $|\gamma|^2 \geq 1$.
Using 
\begin{equation}
Q_i = 4 a_i c_i - b_i^2 \geq 3 a_ic_i 
\end{equation}
now implies 
\begin{equation}
\Nf =   \frac{|\gamma|^2 Q_1 Q_2}{4 c_1 c_2}  \geq \frac{9}{4} a_1 a_2 \ .
\end{equation}
As the minimal value for the $a_i$ is $1$, we find that we only need to consider cases where
\begin{equation}
a_i \leq \frac{4}{9} \Nf \, . 
\end{equation}

If $\Re(\gamma) = 0$ this bound becomes slightly stronger. In this case
\eqref{eq:gamma_conditions} implies 
\begin{equation}
\frac{\Im(\gamma) \sqrt{Q_i}}{2c_i}  \in \Z 
\end{equation}
so that 
\begin{equation}
|\gamma|^2 \geq 4 c_i^2/Q_i
\end{equation}
and we find
\begin{equation} \label{Nbounds}
\Nf \geq 3 c_1 a_2\,, \hspace{1cm} \Nf \geq 3 c_2 a_1 \ .
\end{equation}
As the minimal value for $a_i$ and $c_i$ is $1$, we hence find 
\begin{equation}
a_i \leq \frac{1}{3} \Nf \, 
\end{equation}
in cases where $\Re(\gamma) = 0$. 

In general we can only use the weaker bound which e.g. implies that we can 
find all solutions of tadpole $\leq 30$ by scanning over all values of $b_i \leq c_i \leq a_i \leq 13$. 
The resulting possible values of $\gamma$ and matching pairs of $\K3$ surfaces are recorded in Table \ref{tab:AK-morethan24}. 

For tadpole $\Nf \leq 24 $ the following attractive $\K3$ surfaces appear \cite{Braun:2014ola}:
 \begin{equation}\label{eq:attK3types}
 \begin{aligned}[]
[a,b,c] = [1, 0, 1],[1, 1, 1],[2, 0, 1],[2, 1, 1],[3, 0, 1],[3, 1, 1],[4, 0, 1],[4, 1, 1],\\
  [5, 0, 1],[5, 1, 1],[6, 0, 1],[6, 1, 1],[2, 0,2],[2, 1,2],[2, 2,2],[3, 0,2],\\
  [3, 1,2],[3, 2,2],[4, 0,2],[4, 2,2],[6, 0,2],[6, 2,2],[3, 0,3],[3, 3,3],\\
  [6, 0,3],[6, 3,3],[4, 0,4],[4, 4,4],[5, 0,5],[5, 5,5],[6, 0,6],[6, 6,6],\\
  [7, 7,7],[8, 8,8]\,.
 \end{aligned}
 \end{equation}
For tadpole $\Nf=25$ and $\Nf=26$ there are no solutions, while for $N=27$ the new types
\begin{equation}
[7,1,1],[9,9,9] 
\end{equation}
appear. For $\Nf=28$ we have
\begin{equation}
	[7,0,1],[7,0,7],[8,4,4] \,,
\end{equation}
and for $\Nf=30$
\begin{equation}
	[4,2,4],[8,2,2],[10,10,10] \,,
\end{equation}
without new solutions for $\Nf=29$.

\begin{table}[htbp]
	\begin{center}
		\begin{tabular}{|c|@{}c@{}|c|@{}c@{}|}
			\hline
			$\Nf$ & $[a_1 b_1 c_1]$ & $[a_2 b_2 c_2]$ & $\gamma$  \\
			\hline
			\hline
			30 
			&[10 10 10] & [1 1 1] & $\gamma^{(6)}$ \\
			&[8 2 2] & [4 1 1] & $\pm 2i/\sqrt{15}$  \\
			&[5 5 5] & [2 2 2] & $\gamma^{(6)}$ \\
			&[4 2 4] & [2 1 2] & $\pm 4i/\sqrt{15}$ \\
			&  \multirow{2}{*}{[4 1 1]}&\multirow{2}{*}{[2 1 2]}& $\pm 4i/\sqrt{15}$, \\
			&   & & $\pm(1 + i/\sqrt{15})$ \\
			\hline
			28 
			&[8 4 4] & [2 1 1] & $\pm 2i/\sqrt{7}$ \\
			&[7 0 1] & [7 0 1] & $\pm i/\sqrt{7}$ \\
			&[7 0 7] & [1 0 1] & $\gamma^{(4)}$ \\
			&[7 0 1] & [2 1 1] & $\pm 2 i/\sqrt{7}$  \\
			&[4 2 2] & [4 2 2] & $\pm 2 i/\sqrt{7}$ \\
			&[4 2 2] & [2 1 1] & $\pm 1 \pm i/\sqrt{7}$ \\
			& \multirow{2}{*}{[2 1 1]} & \multirow{2}{*}{[2 1 1]} & $\pm 4i/\sqrt{7}$, \\
			&   &  & $\pm 1 \pm 3i/\sqrt{7}$ \\
			\hline
			27 &[9 9 9] & [1 1 1] & $\gamma^{(6)}$ \\
			&[7 1 1] & [7 1 1] & $\pm 2i/(3\sqrt{3})$ \\
			&[7 1 1] & [1 1 1] &  $\gamma^{(6)}$  \\
			&[3 3 3] & [3 3 3] & $\gamma^{(6)}$ \\
			&[3 3 3] & [1 1 1] & $i \sqrt{3} \gamma^{(6)}$ \\
			& [1 1 1] & [1 1 1] & 3 $\gamma^{(6)}$ \\
			\hline
			24 & [8 8 8] & [1 1 1] & $\gamma^{(6)}$  \\
			& [6 0 6] & [1 0 1] & $\gamma^{(4)}$  \\
			& [6 0 3] & [2 0 1] & $\pm i/\sqrt{2}$ \\
			& [6 0 2] & [3 0 1] & $\pm i / \sqrt{3}$ \\
			& [6 0 2] & [1 1 1] & $\gamma^{(6)}$\\
			& [6 0 1] & [6 0 1] & $ \pm i / \sqrt{6}$ \\
			& [4 4 4] & [2 2 2] & $\gamma^{(6)}$  \\
			& [3 0 3] & [2 0 2] & $\gamma^{(4)}$  \\
			& [3 0 3] & [1 0 1] & $(1+i)\gamma^{(4)}$ \\
			& [3 0 2] & [3 0 2] & $\pm i \sqrt{2/3}$ \\
			& [3 0 1] & [2 2 2] & $\gamma^{(6)}$  \\
			& [2 2 2] & [1 1 1] & $2\gamma^{(6)}$ \\
			& [2 0 1] & [2 0 1] & $\pm 1 \pm i/\sqrt{2}$ \\
			\hline
			23 & [6 1 1] & [6 1 1] & $\pm 2 i/\sqrt{23}$\\
			& [3 1 2] & [3 1 2] & $\pm 4i / \sqrt{23}$  \\
			\hline
			22 & [6 2 2] & [3 1 1] & $\pm 2 i/\sqrt{11}$ \\
			\hline
			21 & [7 7 7] & [1 1 1] & $\gamma^{(6)}$ \\
			& [6 3 3] & [2 1 1] & $\pm 2 i/\sqrt{7}$  \\
			& [1 1 1] & [1 1 1] & $(2 \pm \sqrt{3}i) \gamma^{(6)}$ \\
			\hline
		\end{tabular}
		\begin{tabular}{|c|c|c|@{}c@{}|}
			\hline
			$\Nf$ & $[a_1 b_1 c_1]$ & $[a_2 b_2 c_2]$ & $\gamma$  \\
			\hline
			\hline
			20 & [5 0 5] & [1 0 1] & $\gamma^{(4)}$  \\
			& [5 0 1] & [5 0 1] & $\pm i/\sqrt{5}$  \\
			& [3 2 2] & [3 2 2] & $\pm 2 i/\sqrt{5}$  \\
			& [1 0 1] & [1 0 1] & $(1 \pm 2i) \gamma^{(4)}$  \\
			\hline 
			19 & [5 1 1] & [5 1 1] & $\pm 2 i /\sqrt{19}$  \\
			\hline
			18 & [6 6 6] & [1 1 1] & $\gamma^{(6)}$ \\
			& [3 3 3] & [2 2 2] & $\gamma^{(6)}$\\
			\hline
			16 & [4 0 4] & [1 0 1] & $\gamma^{(4)}$  \\
			& [4 0 2] & [2 0 1] & $\pm i/\sqrt{2}$  \\
			& [4 0 1] & [4 0 1] & $\pm i/2$  \\
			& [4 0 1] & [1 0 1] & $\gamma^{(4)}$  \\
			& [2 0 2] & [2 0 2] & $\gamma^{(4)}$  \\
			& [2 0 2] & [1 0 1] & $(1+i)\gamma^{(4)}$ \\
			& [2 0 1] & [2 0 1] & $\pm 1$  \\
			& [1 0 1] & [1 0 1] & $2 \gamma^{(4)}$  \\ 
			\hline
			15 & [5 5 5] & [1 1 1] & $\gamma^{(6)}$ \\
			& [4 1 1] & [4 1 1] & $\pm 2 i/\sqrt{15}$ \\
			& [2 1 2] & [2 1 2] & $\pm 4 i /\sqrt{15}$  \\ 
			\hline
			14 & [4 2 2] & [2 1 1] & $\pm 2i / \sqrt{7}$  \\
			& [2 1 1] & [2 1 1] & $\pm 1 \pm i /\sqrt{7}$  \\
			\hline
			12 & [4 4 4] & [1 1 1] & $\gamma^{(6)}$  \\
			& [3 0 3] & [1 0 1] & $\gamma^{(4)}$  \\
			& [3 0 1] & [3 0 1] & $\pm i/\sqrt{3}$ \\
			& [3 0 1] & [1 1 1] & $\gamma^{(6)}$  \\
			& [2 2 2] & [2 2 2] & $\gamma^{(6)}$  \\
			& [1 1 1] & [1 1 1] & $2 \gamma^{(6)}$ \\
			\hline
			11 & [3 1 1] & [3 1 1] & $\pm 2 i /\sqrt{11}$  \\
			\hline 
			9 & [3 3 3] & [1 1 1] & $\gamma^{(6)}$ \\
			& [1 1 1] & [1 1 1] & $\sqrt{3}i \gamma^{(6)}$  \\
			\hline
			8 & [2 0 2] & [1 0 1] & $\gamma^{(4)}$  \\
			& [2 0 1] & [2 0 1] & $\pm i/\sqrt{2}$  \\
			& [1 0 1] & [1 0 1] & $(1+i)\gamma^{(4)}$  \\
			\hline
			7 & [2 1 1] & [2 1 1] & $\pm 2 i /\sqrt{7}$  \\
			\hline 
			6 & [2 2 2] & [1 1 1] & $\gamma^{(6)}$  \\
			\hline
			4 & [1 0 1] & [1 0 1] & $\gamma^{(4)}$  \\
			\hline
			3 & [1 1 1] & [1 1 1] & $\gamma^{(6)}$  \\
			\hline
			\multicolumn{4}{c}{}  \\
			\multicolumn{4}{c}{}  \\
			\multicolumn{4}{c}{}  \\
		\end{tabular}
		\caption{\label{tab:AK-morethan24}All solutions with $\Nf \leq 30$, table partly reproduced and extended from \cite{Braun:2014ola}. The notation $[a,b,c]$ for the $\K3$ surfaces is introduced in \eqref{abc} and the constant  $\gamma$ in the four-form flux in \eqref{eq:fluxpuretype0}. Here $\gamma^{(6)}= \frac{2i}{\sqrt{3}} e^{2\pi i k/6}$, with $k=1,..,6$ and $\gamma^{(4)}= e^{2\pi i k/4}$ with $k=1,..,4$.}
 \end{center}
\end{table}

\subsection{F-theory lifts and the Kneser-Nishiyama method}\label{sect:FliftandKN}

Having found all flux solutions in M-theory, we are now ready to construct all F-theory lifts. As the fluxes \eqref{eq:fluxquantomega} are always of the type that lifts to F-theory, we only need to choose an elliptic fibration on either $S_1$ or $S_2$. After this choice, we can read off the gauge group from the frame lattice. 
The set of possible gauge groups can hence be inferred by studying which frame lattices can occur for any of the attractive $\K3$ surfaces appearing in \eqref{eq:attK3types}. 

Let us simply denote the attractive $\K3$ surface that is equipped with an elliptic fibration by $S$. As explained in section \ref{sec:MFK3}, an elliptic fibration on $S$ is specified by a primitive embedding $U \hookrightarrow Pic(S)$. Once such an embedding is specified we may write
\begin{equation}\label{eq:pic_decompose}
Pic(S) = U \oplus W 
\end{equation}
where $W$ is the frame lattice of the elliptic fibration which for attactive $\K3$ surfaces has signature $(0,18)$. 

Instead of trying to determine all embeddings $U \hookrightarrow Pic(S)$ modulo isomorphism, we use the Kneser-Nishiyama method
which greatly simplifies the problem and gives us direct access to the frame lattices $W$ which occur for various elliptic fibrations.\footnote{As discussed 
above, this determines the singular fibres and Mordell-Weil group. It does not however uniquely fix the elliptic fibration up to isomorphism of $S$. In other 
words, the same collections of singular fibres and Mordell-Weil groups might appear for several elliptic fibrations on the same attractive $\K3$ surface, without 
these being isomorphic as complex surfaces; see \cite{Braun:2013yya} for a detailed discussion.}

For a given $Pic(S)$, all possible $W$ can be found as follows. By theorem \ref{thm:existence_of_emb} (see Appendix \ref{app:primitive_embeddings}),
any $T_S$ can be embedded (not necessary uniquely) into the root lattice $E_8$, and we define 
\begin{equation}
T_0 := T_S^\perp \subset E_8 
\end{equation}
which is lattice of signature $(6,0)$. 

Any $W$ which appears in \eqref{eq:pic_decompose} must be such that $q(W) = q\left(Pic(S)\right) = - q(T_S)$ by theorem \ref{thm:orthcomp}.  
Hence we also have that $q(W[-1]) = -q(T_0)$ and there must exist an even self-dual lattice $N$ of signature $(24,0)$ in which both $W$ and 
$T_0$ are each primitively embedded such that they are orthogonal complements. Even self-dual lattices of signature $(24,0)$ are classified and are known
as the $24$ Niemeier lattices $N_I$; see Appendix \ref{app:niemeier} for a brief review. Any $W$ which appears in \eqref{eq:pic_decompose} must hence be such that one of the $N_I$ is an overlattice of 
$W \oplus T_0$ with $W$ and $T_0$ each primitively embedded and mutually orthogonal, 
\begin{equation}
 N_I \supseteq W \oplus T_0 \, .
\end{equation}
As this holds for every choice of $T_0$, we can proceed without loss of generality by picking one $T_0$ among the different possibilities. We can then determine all $W$'s by finding all the primitive embeddings of our chosen $T_0$ into all of the Niemeier lattices $N_I$:
\begin{equation}
\{W\} \cong \bigcup_I \{i: T_0 \hookrightarrow N_I |\, i\,\, \mbox{primitive}\}\, ,
\end{equation}
where it is understood that we delete duplicates on the right-hand side. 

Phrased in this way, the determination of all possible frame lattices 
becomes feasible. We have given a list of the Niemeier lattices in Appendix \ref{app:niemeier} and in Appendix \ref{app:embd_in_Niemeier} we have collected some useful results which allow us to classify 
embeddings of root lattices in them. All of the Niemeier lattices except one (the Leech lattice) are constructed using ADE root lattices, reviewed in Appendix \ref{app:root_lattices}. In the cases of main interest to us, $T_0$ contains a root lattice of rank at least $4$, which 
greatly simplifies classifying primitive embeddings. We will also consider a few cases with $N_\text{flux} \geq 25$, which are not necessarily subject to this simplification.

Also note that the above allows us to show that any attractive $\K3$ surface admits an elliptic fibration. First note that fixing the lattice $T_S$ uniquely determins the embedding
of $T_S$ into $\Lambda^{3,19}$ by Theorem \ref{thm:uniqueness_of_emb}, so that $Pic(S)$ is uniquely determined as well. For any $T_S$ of rank two, there always exists a primitive embedding into $E_8$ by Theorem \eqref{thm:existence_of_emb}, so that we can always find a suitable $T_0$. The same theorem then guarantees that $T_0$ embeds into some Niemeier lattice. This also follows from our construction of $T_0$ and the fact that one of the Niemeier lattices is $E_8^{3}$. The orthogonal complement then gives us a suitable frame lattice $W$ on $S$. It then follows that $W \oplus U \oplus T_S = Pic(S) \oplus T_S$ has $\Lambda^{3,19}$ as an overlattice, showing the existence of an elliptic fibration.

\section{Method and Results}
\label{sec:results}

In this section we explain how to use the Kneser-Nishiyama method to examine the frame lattices discussed above for every elliptically fibered $\K3$ allowing to satisfy the tadpole bound $N_\text{flux} \leq 24$, and report on the results of a systematic scan regarding the properties of their root sublattices in relation to the tadpole conjecture. We show that the conjecture holds for this type of flux vacua. We also consider the cases corresponding to $25 \leq N_{\text{flux}} \leq 30$ to understand the behaviour of symmetry enhancements above the tadpole bound; we find that no frame lattice without roots exists for $\Nf < 30$. At $\Nf = 30$, however, we show that there is indeed a frame lattice without roots by giving an explicit construction.
\subsection{Finding \texorpdfstring{$T_0$}{T0}}

The first step in the Kneser-Nishiyama method is to find a lattice $T_0$ for every $T_S$ of a suitable form; namely, what we look for is a lattice whose generating vectors are as short as possible (e.g. with many roots), facilitating its embedding into the Niemeier lattices. To this end we can proceed by directly computing orthogonal complements of $T_S$ for various embeddings into $E_8$ and compiling lists of the resulting $T_0$'s, picking the one satisfying our requirements. The computation of all possible $T_0$'s is, however, amenable to exact algorithmic solutions implemented for example in \texttt{SAGEMATH} --- it is nothing more than the computation of a lattice genus.\footnote{For even lattices, a genus is the set of all lattices of a given rank sharing a given discriminant quadratic form. Note that the set of frame lattices corresponding to a given $\K3$ surface is also a genus, but its computation is generically not feasible with direct algorithmic tools. This is largely due to the high rank of frame lattices, in this case 18.} For illustrative purposes, in the following we describe a strategy for the former procedure, whose results can be cross-checked with those obtained with \texttt{SAGEMATH}.

The computation of $T_0$'s proceeds as follows. First we need to generate an arbitrary primitive embedding of $T_S \equiv [a~b~c]$ into $E_8$. This is achieved by picking a random primitive vector $t_1 \in E_8$ with norm $t_1^2 = 2a$ and then another random primitive vector $t_2 \in E_8$ with norm $t_2^2 = 2c$ and $t_1 \cdot t_2 = b$; primitivity can be checked by computing the vectors in the dual of the lattice generated by $t_{1,2}$ and determining that none lie inside $E_8$ except those in the original lattice (more precisely one needs only to check this for vectors with even norm.)  

Next we look for a set of six linearly independent vectors in $E_8$ orthogonal to $t_1$ and $t_2$, which generically generate a sublattice $\tilde T_0$ of $T_0$. In practice, we construct $\tilde T_0$ by looking for vectors orthogonal to $T_S$ with the smallest possible norm, starting with roots, and then progressively looking for vectors with greater norm if required. All overlattices of $\tilde T_0$ can then be constructed by appropriately replacing some of its generators with vectors in the dual lattice $\tilde T_0^*$. The lattice $T_0$ corresponds to the case in which the resulting overlattice is embedded in $E_8$ and its determinant is equal to $\text{det}(T_S) = 4ac - b^2$.

Out of the complete lists of $T_0$'s for each $T_S$, we have picked those shown in Table \ref{tab:Dynkins}, where they are represented using Dynkin diagrams and generalizations thereof explained in the caption. For example, for the lattice $\begin{bmatrix}4&0&4\\\end{bmatrix}$ we chose
\begin{equation}
	T_0 = 
	\begin{pmatrix}
	2&-1&&&&\\
	-1&2&+1&-1&&\\
	&+1&4&&-1&\\
	&-1&&4&-1&\\
	&&-1&-1&2&-1\\
	&&&&-1&2
	\end{pmatrix}~~~\simeq~~~
	\begin{tikzpicture}
	\draw (0,0)--(2,0);
	\draw (0.5,0)--(1,0.5);
	\draw (1.5,0)--(1,0.5);
	\draw (0.75,-0.1)--(0.75,0.1);
	\draw[fill=white] (0,0) circle(0.1);
	\draw[fill=white] (0.5,0) circle(0.1);
	\draw[fill=black] (1,0) circle(0.1);
	\draw[fill=white] (1.5,0) circle(0.1);
	\draw[fill=white] (2,0) circle(0.1);
	\draw[fill=black] (1,0.5) circle(0.1);
	\end{tikzpicture}\,,
\end{equation}
which exhibits the most complicated structure arising with $N_\text{flux} \leq 24$. Indeed, out of the 34 lattices in this class, 11 are root lattices, 14 have one vector with norm 4, six have two vectors with norm 4, one has one vector with norm 6 and two have two vectors with norm 6; only the one shown above requires a positive inner product. As we will see, the problem of studying frame lattices associated to each $T_S$ is well under control for all but the cases in which the $T_0$ has two vectors with norm 6. For these exceptional cases we are able to obtain a comprehensive amount of data, but cannot guarantee an exhaustive count of the possible frame lattices. Since we are particularly interested in the question of whether or not there exist frame lattices with no roots, which requires exhaustivity, we will employ a complementary algorithm designed to answer this specific question. 

We have also computed eight $T_0$'s corresponding to the possible $T_S$'s admitting $25 \leq N_\text{flux} \leq 30$ in order to study how non-Abelian symmetry enhancement behaves above the tadpole bound but still close to it. As is easily seen from Table \ref{tab:Dynkins}, some of these lattices take on more complicated forms and we treat them in a similar way to the two exceptional lattices mentioned above.

\begin{table}[htbp]
	\begin{center}
		\begin{tabular}{|c|c|c|c|c|c|}
			\hline
			$\begin{smallmatrix}1&0&1\\\end{smallmatrix}$& 
			$\begin{smallmatrix}1&1&1\\\end{smallmatrix}$& 
			$\begin{smallmatrix}2&0&1\\\end{smallmatrix}$& 
			$\begin{smallmatrix}2&0&2\\\end{smallmatrix}$& 
			$\begin{smallmatrix}2&1&1\\\end{smallmatrix}$& 
			$\begin{smallmatrix}2&1&2\\\end{smallmatrix}$
			\\\hline
			\rule{0pt}{30pt}
			\begin{tikzpicture}
				\draw (0,0)--(2,0);
				\draw (0.5,0)--(0.5,0.5);
				\draw[fill=white] (0,0) circle(0.1);
				\draw[fill=white] (0.5,0) circle(0.1);
				\draw[fill=white] (1,0) circle(0.1);
				\draw[fill=white] (1.5,0) circle(0.1);
				\draw[fill=white] (2,0) circle(0.1);
				\draw[fill=white] (0.5,0.5) circle(0.1);
			\end{tikzpicture}
			&
			\begin{tikzpicture}
				\draw (0,0)--(2,0);
				\draw (1,0)--(1,0.5);
				\draw[fill=white] (0,0) circle(0.1);
				\draw[fill=white] (0.5,0) circle(0.1);
				\draw[fill=white] (1,0) circle(0.1);
				\draw[fill=white] (1.5,0) circle(0.1);
				\draw[fill=white] (2,0) circle(0.1);
				\draw[fill=white] (1,0.5) circle(0.1);
			\end{tikzpicture}
			&
			\begin{tikzpicture}
				\draw (0,0)--(1.5,0);
				\draw (0.5,0)--(0.5,0.5);
				\draw[fill=white] (0,0) circle(0.1);
				\draw[fill=white] (0.5,0) circle(0.1);
				\draw[fill=white] (1,0) circle(0.1);
				\draw[fill=white] (1.5,0) circle(0.1);
				\draw[fill=white] (2,0) circle(0.1);
				\draw[fill=white] (0.5,0.5) circle(0.1);
			\end{tikzpicture}
			&
			\begin{tikzpicture}
				\draw (0,0)--(0.5,0);
				\draw (0,0)--(0,0.5);
				\draw (1,0)--(2,0);
				\draw[fill=white] (0,0) circle(0.1);
				\draw[fill=white] (0.5,0) circle(0.1);
				\draw[fill=white] (1,0) circle(0.1);
				\draw[fill=white] (1.5,0) circle(0.1);
				\draw[fill=white] (2,0) circle(0.1);
				\draw[fill=white] (0,0.5) circle(0.1);
			\end{tikzpicture}
			&
			\begin{tikzpicture}
				\draw (0,0)--(2,0);
				\draw (0,0)--(0,0.5);
				\draw[fill=white] (0,0) circle(0.1);
				\draw[fill=white] (0.5,0) circle(0.1);
				\draw[fill=white] (1,0) circle(0.1);
				\draw[fill=white] (1.5,0) circle(0.1);
				\draw[fill=white] (2,0) circle(0.1);
				\draw[fill=white] (0,0.5) circle(0.1);
			\end{tikzpicture}
			& 
			\begin{tikzpicture}
				\draw (0,0)--(2,0);
				\draw (1,0)--(1,0.5);
				\draw[fill=white] (0,0) circle(0.1);
				\draw[fill=white] (0.5,0) circle(0.1);
				\draw[fill=white] (1,0) circle(0.1);
				\draw[fill=white] (1.5,0) circle(0.1);
				\draw[fill=white] (2,0) circle(0.1);
				\draw[fill=black] (1,0.5) circle(0.1);
			\end{tikzpicture}
			\\
			\hline\hline
			$\begin{smallmatrix}2&2&2\\\end{smallmatrix}$& 
			$\begin{smallmatrix}3&0&1\\\end{smallmatrix}$& 
			$\begin{smallmatrix}3&0&2\\\end{smallmatrix}$& 
			$\begin{smallmatrix}3&0&3\\\end{smallmatrix}$& 
			$\begin{smallmatrix}3&1&1\\\end{smallmatrix}$&
			$\begin{smallmatrix}3&1&2\\\end{smallmatrix}$  
			\\\hline
			\rule{0pt}{30pt}
			\begin{tikzpicture}
				\draw (0,0)--(1,0);
				\draw (1.5,0)--(2,0);
				\draw (0.5,0)--(0.5,0.5);
				\draw[fill=white] (0,0) circle(0.1);
				\draw[fill=white] (0.5,0) circle(0.1);
				\draw[fill=white] (1,0) circle(0.1);
				\draw[fill=white] (1.5,0) circle(0.1);
				\draw[fill=white] (2,0) circle(0.1);
				\draw[fill=white] (0.5,0.5) circle(0.1);
			\end{tikzpicture}
			& 
			\begin{tikzpicture}
				\draw (0,0)--(2,0);
				\draw[fill=white] (0,0) circle(0.1);
				\draw[fill=white] (0.5,0) circle(0.1);
				\draw[fill=white] (1,0) circle(0.1);
				\draw[fill=white] (1.5,0) circle(0.1);
				\draw[fill=white] (2,0) circle(0.1);
				\draw[fill=white] (1,0.5) circle(0.1);
			\end{tikzpicture}
			&
			\begin{tikzpicture}
				\draw (0,0)--(2,0);
				\draw[fill=white] (0,0) circle(0.1);
				\draw[fill=white] (0.5,0) circle(0.1);
				\draw[fill=white] (1,0) circle(0.1);
				\draw[fill=white] (1.5,0) circle(0.1);
				\draw[fill=white] (2,0) circle(0.1);
				\draw[fill=black] (1,0.5) circle(0.1);
			\end{tikzpicture}
			&
			\begin{tikzpicture}
				\draw (0,0)--(0.5,0);
				\draw (1.5,0)--(2,0);
				\draw[fill=white] (0,0) circle(0.1);
				\draw[fill=white] (0.5,0) circle(0.1);
				\draw[fill=white] (1,0) circle(0.1);
				\draw[fill=white] (1.5,0) circle(0.1);
				\draw[fill=white] (2,0) circle(0.1);
				\draw[fill=white] (1,0.5) circle(0.1);
			\end{tikzpicture}
			&
			\begin{tikzpicture}
				\draw (0,0)--(2,0);
				\draw (1,0)--(1,0.5);
				\draw[fill=black] (0,0) circle(0.1);
				\draw[fill=white] (0.5,0) circle(0.1);
				\draw[fill=white] (1,0) circle(0.1);
				\draw[fill=white] (1.5,0) circle(0.1);
				\draw[fill=white] (2,0) circle(0.1);
				\draw[fill=white] (1,0.5) circle(0.1);
			\end{tikzpicture}
			&
			\begin{tikzpicture}
				\draw (0,0)--(2,0);
				\draw (1,0)--(1,0.5);
				\draw[fill=white] (0,0) circle(0.1);
				\draw[fill=white] (0.5,0) circle(0.1);
				\draw[fill=white] (1,0) circle(0.1);
				\draw[fill=black] (1.5,0) circle(0.1);
				\draw[fill=white] (2,0) circle(0.1);
				\draw[fill=white] (1,0.5) circle(0.1);
			\end{tikzpicture}			
			\\
			\hline\hline
			$\begin{smallmatrix}3&2&2\\\end{smallmatrix}$& 
			$\begin{smallmatrix}3&3&3\\\end{smallmatrix}$& 
			$\begin{smallmatrix}4&0&1\\\end{smallmatrix}$& 
			$\begin{smallmatrix}4&0&2\\\end{smallmatrix}$& 
			$\begin{smallmatrix}4&0&4\\\end{smallmatrix}$&
			$\begin{smallmatrix}4&1&1\\\end{smallmatrix}$
			\\\hline 
			\rule{0pt}{30pt}
			\begin{tikzpicture}
				\draw (0.5,0)--(2,0);
				\draw[fill=white] (0,0) circle(0.1);
				\draw[fill=white] (0.5,0) circle(0.1);
				\draw[fill=white] (1,0) circle(0.1);
				\draw[fill=white] (1.5,0) circle(0.1);
				\draw[fill=white] (2,0) circle(0.1);
				\draw[fill=white] (0,0.5) circle(0.1);
			\end{tikzpicture}
			&
			\begin{tikzpicture}
				\draw (0,0)--(2,0);
				\draw (1,0)--(1,0.5);
				\draw[fill=black] (0,0) circle(0.1);
				\draw[fill=white] (0.5,0) circle(0.1);
				\draw[fill=white] (1,0) circle(0.1);
				\draw[fill=white] (1.5,0) circle(0.1);
				\draw[fill=black] (2,0) circle(0.1);
				\draw[fill=white] (1,0.5) circle(0.1);
			\end{tikzpicture}
			&
			\begin{tikzpicture}
				\draw (0,0)--(2,0);
				\draw (0.5,0)--(0.5,0.5);
				\draw (0.5,0)--(0.5,0.5);
				\draw[fill=white] (0,0) circle(0.1);
				\draw[fill=white] (0.5,0) circle(0.1);
				\draw[fill=white] (1,0) circle(0.1);
				\draw[fill=white] (1.5,0) circle(0.1);
				\draw[fill=white] (2,0) circle(0.1);
				\draw[fill=black] (0.5,0.5) circle(0.1);
			\end{tikzpicture}
			&
			\begin{tikzpicture}
				\draw (0,0)--(1.5,0);
				\draw (0.5,0)--(0.5,0.5);
				\draw (1,0)--(1,0.5);
				\draw[fill=white] (0,0) circle(0.1);
				\draw[fill=black] (0.5,0) circle(0.1);
				\draw[fill=white] (1,0) circle(0.1);
				\draw[fill=white] (1.5,0) circle(0.1);
				\draw[fill=white] (1,0.5) circle(0.1);
				\draw[fill=white] (0.5,0.5) circle(0.1);
			\end{tikzpicture}
			&
			\begin{tikzpicture}
				\draw (0,0)--(2,0);
				\draw (0.5,0)--(1,0.5);
				\draw (1.5,0)--(1,0.5);
				\draw (0.75,-0.1)--(0.75,0.1);
				\draw[fill=white] (0,0) circle(0.1);
				\draw[fill=white] (0.5,0) circle(0.1);
				\draw[fill=black] (1,0) circle(0.1);
				\draw[fill=white] (1.5,0) circle(0.1);
				\draw[fill=white] (2,0) circle(0.1);
				\draw[fill=black] (1,0.5) circle(0.1);
			\end{tikzpicture}
			&
			\begin{tikzpicture}
				\draw (0.5,0)--(2,0);
				\draw (0,0)--(0,0.5);
				\draw[fill=white] (0,0) circle(0.1);
				\draw[fill=white] (0.5,0) circle(0.1);
				\draw[fill=white] (1,0) circle(0.1);
				\draw[fill=white] (1.5,0) circle(0.1);
				\draw[fill=white] (2,0) circle(0.1);
				\draw[fill=white] (0,0.5) circle(0.1);
			\end{tikzpicture}
			\\
			\hline\hline
			$\begin{smallmatrix}4&2&2\\\end{smallmatrix}$& 
			$\begin{smallmatrix}4&4&4\\\end{smallmatrix}$& 
			$\begin{smallmatrix}5&0&1\\\end{smallmatrix}$& 
			$\begin{smallmatrix}5&0&5\\\end{smallmatrix}$& 
			$\begin{smallmatrix}5&1&1\\\end{smallmatrix}$&
			$\begin{smallmatrix}5&5&5\\\end{smallmatrix}$
			\\\hline 
			\rule{0pt}{30pt}
			\begin{tikzpicture}
				\draw (0,0)--(2,0);
				\draw (0.5,0)--(0.5,0.5);
				\draw[fill=white] (0,0) circle(0.1);
				\draw[fill=white] (0.5,0) circle(0.1);
				\draw[fill=black] (1,0) circle(0.1);
				\draw[fill=white] (1.5,0) circle(0.1);
				\draw[fill=white] (2,0) circle(0.1);
				\draw[fill=white] (0.5,0.5) circle(0.1);
			\end{tikzpicture}
			&
			\begin{tikzpicture}
				\draw (0,0)--(2,0);
				\draw (0.5,0)--(1,0.5);
				\draw (1.5,0)--(1,0.5);
				\draw[fill=white] (0,0) circle(0.1);
				\draw[fill=white] (0.5,0) circle(0.1);
				\draw[fill=black] (1,0) circle(0.1);
				\draw[fill=white] (1.5,0) circle(0.1);
				\draw[fill=white] (2,0) circle(0.1);
				\draw[fill=black] (1,0.5) circle(0.1);
			\end{tikzpicture}
			&
			\begin{tikzpicture}
				\draw (0,0)--(2,0);
				\draw (0.5,0)--(0.5,0.5);
				\draw[fill=white] (0,0) circle(0.1);
				\draw[fill=white] (0.5,0) circle(0.1);
				\draw[fill=white] (1,0) circle(0.1);
				\draw[fill=black] (1.5,0) circle(0.1);
				\draw[fill=white] (2,0) circle(0.1);
				\draw[fill=white] (0.5,0.5) circle(0.1);
			\end{tikzpicture}
			&
			\begin{tikzpicture}
				\draw (0,0)--(0.5,0);
				\draw (1,0)--(2,0);
				\draw (0,0)--(0,0.5);
				\draw[fill=white] (0,0) circle(0.1);
				\draw[fill=white] (0.5,0) circle(0.1);
				\draw[fill=black] (1,0) circle(0.1);
				\draw[fill=white] (1.5,0) circle(0.1);
				\draw[fill=white] (2,0) circle(0.1);
				\draw[fill=black] (0,0.5) circle(0.1);
			\end{tikzpicture}
			&
			\begin{tikzpicture}
				\draw (0,0)--(2,0);
				\draw (0,0)--(0,0.5);
				\draw[fill=white] (0,0) circle(0.1);
				\draw[fill=white] (0.5,0) circle(0.1);
				\draw[fill=white] (1,0) circle(0.1);
				\draw[fill=white] (1.5,0) circle(0.1);
				\draw[fill=white] (2,0) circle(0.1);
				\draw[fill=black] (0,0.5) circle(0.1);
			\end{tikzpicture}
			&
			\begin{tikzpicture}
				\draw (0,0)--(1,0);
				\draw (1.5,0)--(2,0);
				\draw (0.5,0)--(1,0.5);
				\draw (1,0)--(1,0.5);
				\draw[fill=white] (0,0) circle(0.1);
				\draw[fill=white] (0.5,0) circle(0.1);
				\draw[fill=black] (1,0) circle(0.1);
				\draw[fill=white] (1.5,0) circle(0.1);
				\draw[fill=white] (2,0) circle(0.1);
				\draw[fill=black] (1,0.5) circle(0.1);
			\end{tikzpicture}
			\\
			\hline\hline
			$\begin{smallmatrix}6&0&1\\\end{smallmatrix}$& 
			$\begin{smallmatrix}6&0&2\\\end{smallmatrix}$& 
			$\begin{smallmatrix}6&0&3\\\end{smallmatrix}$& 
			$\begin{smallmatrix}6&0&6\\\end{smallmatrix}$& 
			$\begin{smallmatrix}6&1&1\\\end{smallmatrix} \simeq \begin{smallmatrix}3&1&2\\\end{smallmatrix}$ &
			$\begin{smallmatrix}6&2&2\\\end{smallmatrix}$
			\\\hline 
			\rule{0pt}{30pt}
			\begin{tikzpicture}
				\draw (0,0)--(0.5,0);
				\draw (1,0)--(2,0);
				\draw[fill=white] (0,0) circle(0.1);
				\draw[fill=white] (0.5,0) circle(0.1);
				\draw[fill=white] (1,0) circle(0.1);
				\draw[fill=white] (1.5,0) circle(0.1);
				\draw[fill=white] (2,0) circle(0.1);
				\draw[fill=white] (0,0.5) circle(0.1);
			\end{tikzpicture}
			&
			\begin{tikzpicture}
				\draw (0,0)--(0.5,0);
				\draw (1,0)--(2,0);
				\draw[fill=white] (0,0) circle(0.1);
				\draw[fill=white] (0.5,0) circle(0.1);
				\draw[fill=white] (1,0) circle(0.1);
				\draw[fill=white] (1.5,0) circle(0.1);
				\draw[fill=white] (2,0) circle(0.1);
				\draw[fill=black] (0,0.5) circle(0.1);
			\end{tikzpicture}
			&
			\begin{tikzpicture}
				\draw (0,0)--(0.5,0);
				\draw (1,0)--(2,0);
				\draw[fill=white] (0,0) circle(0.1);
				\draw[fill=white] (0.5,0) circle(0.1);
				\draw[fill=white] (1,0) circle(0.1);
				\draw[fill=black] (1.5,0) circle(0.1);
				\draw[fill=white] (2,0) circle(0.1);
				\draw[fill=white] (0,0.5) circle(0.1);
			\end{tikzpicture}
			&
			\begin{tikzpicture}
				\draw (0,0)--(0.5,0);
				\draw (1.5,0)--(2,0);
				\draw[fill=white] (0,0) circle(0.1);
				\draw[fill=white] (0.5,0) circle(0.1);
				\draw[fill=black] (1,0) circle(0.1);
				\draw[fill=white] (1.5,0) circle(0.1);
				\draw[fill=white] (2,0) circle(0.1);
				\draw[fill=black] (1,0.5) circle(0.1);
			\end{tikzpicture}
			&
			\begin{tikzpicture}
				\draw (0,0)--(2,0);
				\draw (1,0)--(1,0.5);
				\draw[fill=white] (0,0) circle(0.1);
				\draw[fill=white] (0.5,0) circle(0.1);
				\draw[fill=white] (1,0) circle(0.1);
				\draw[fill=black] (1.5,0) circle(0.1);
				\draw[fill=white] (2,0) circle(0.1);
				\draw[fill=white] (1,0.5) circle(0.1);
			\end{tikzpicture}
			&
			\begin{tikzpicture}
				\draw (0,0)--(1.5,0);
				\draw (0.5,0)--(0.5,0.5);
				\draw (0.5,0)--(1,0.5);
				\draw[fill=white] (0,0) circle(0.1);
				\draw[fill=black] (0.5,0) circle(0.1);
				\draw[fill=white] (1,0) circle(0.1);
				\draw[fill=white] (1.5,0) circle(0.1);
				\draw[fill=white] (1,0.5) circle(0.1);
				\draw[fill=white] (0.5,0.5) circle(0.1);
			\end{tikzpicture}
			\\
			\hline\hline
			$\begin{smallmatrix}6&3&3\\\end{smallmatrix}$& 
			$\begin{smallmatrix}6&6&6\\\end{smallmatrix}$& 
			$\begin{smallmatrix}7&7&7\\\end{smallmatrix}$& 
			$\begin{smallmatrix}8&8&8\\\end{smallmatrix}$& 
			$*\begin{smallmatrix}7&1&1\\\end{smallmatrix}*$&
			$*\begin{smallmatrix}9&9&9\\\end{smallmatrix}* $
			\\\hline 
			\rule{0pt}{30pt}
			\begin{tikzpicture}
				\draw (0,0)--(0,0.5);
				\draw (0.5,0)--(1,0);
				\draw (1.5,0)--(2,0);
				\draw[fill=white] (0,0) circle(0.1);
				\draw[fill=white] (0.5,0) circle(0.1);
				\draw[fill=white] (1,0) circle(0.1);
				\draw[fill=white] (1.5,0) circle(0.1);
				\draw[fill=white] (2,0) circle(0.1);
				\draw[fill=black] (0,0.5) circle(0.1);
			\end{tikzpicture}
			&
			\begin{tikzpicture}
				\draw (0.5,0)--(0.5,0.5);
				\draw (0,0)--(1,0);
				\draw (1.5,0)--(2,0);
				\draw[fill=white] (0,0) circle(0.1);
				\draw[fill=black] plot [only marks, mark = square*] coordinates {(0.5,0)};
				\draw[fill=white] (1,0) circle(0.1);
				\draw[fill=white] (1.5,0) circle(0.1);
				\draw[fill=white] (2,0) circle(0.1);
				\draw[fill=white] (0.5,0.5) circle(0.1);
			\end{tikzpicture}
			&
			\begin{tikzpicture}
				\draw (0,0)--(2,0);
				\draw (0.5,0)--(1,0.5);
				\draw (1,0)--(1,0.5);
				\draw (1.5,0)--(1,0.5);
				\draw[fill=white] (0,0) circle(0.1);
				\draw[fill=white] (0.5,0) circle(0.1);
				\draw[fill=black] plot [only marks, mark = square*] coordinates {(1,0)};
				\draw[fill=white] (1.5,0) circle(0.1);
				\draw[fill=white] (2,0) circle(0.1);
				\draw[fill=black] plot [only marks, mark = square*] coordinates {(1,0.5)};
			\end{tikzpicture}
			&
			\begin{tikzpicture}
				\draw (0,0)--(1,0);
				\draw (1.5,0)--(2,0);
				\draw (0.5,0)--(1,0.5);
				\draw (0.95,0)--(0.95,0.5);
				\draw (1.05,0)--(1.05,0.5);
				\draw[fill=white] (0,0) circle(0.1);
				\draw[fill=white] (0.5,0) circle(0.1);
				\draw[fill=black] plot [only marks, mark = square*] coordinates {(1,0)};
				\draw[fill=white] (1.5,0) circle(0.1);
				\draw[fill=white] (2,0) circle(0.1);
				\draw[fill=black] plot [only marks, mark = square*] coordinates {(1,0.5)};
			\end{tikzpicture}
			&
			\begin{tikzpicture}
				\draw (0,0)--(2,0);
				\draw (0,0)--(0,0.5);
				\draw[fill=white] (0,0) circle(0.1);
				\draw[fill=white] (0.5,0) circle(0.1);
				\draw[fill=white] (1,0) circle(0.1);
				\draw[fill=black] (1.5,0) circle(0.1);
				\draw[fill=white] (2,0) circle(0.1);
				\draw[fill=white] (0,0.5) circle(0.1);
			\end{tikzpicture}
			&
			\begin{tikzpicture}
				\draw (0,0)--(1.5,0);
				\draw (0.5,0.5)--(1,0.5);
				\draw (0.5,0)--(0.5,0.5);
				\draw (0.5,0.5)--(1,0);
				\draw[fill=white] (0,0) circle(0.1);
				\draw[fill=black] (0.5,0) circle(0.1);
				\draw[fill=black] (1,0) circle(0.1);
				\draw[fill=white] (1.5,0) circle(0.1);
				\draw[fill=white] (1,0.5) circle(0.1);
				\draw[fill=black] (0.5,0.5) circle(0.1);
			\end{tikzpicture}
			\\\hline\hline
			$*\begin{smallmatrix}7&0&1\\\end{smallmatrix}*$& 
			$*\begin{smallmatrix}7&0&7\\\end{smallmatrix}*$& 
			$*\begin{smallmatrix}8&4&4\\\end{smallmatrix}*$& 
			$*\begin{smallmatrix}4&2&4\\\end{smallmatrix}*$& 
			$*\begin{smallmatrix}8&2&2\\\end{smallmatrix}*$&
			$*\begin{smallmatrix}10&10&10\\\end{smallmatrix}*$
			\\\hline 
			\rule{0pt}{30pt}
			\begin{tikzpicture}
				\draw (0,0)--(1.5,0);
				\draw (0.5,0)--(0.5,0.5);
				\draw (0.5,0)--(0.5,0.5);
				\draw[fill=white] (0,0) circle(0.1);
				\draw[fill=white] (0.5,0) circle(0.1);
				\draw[fill=white] (1,0) circle(0.1);
				\draw[fill=white] (1.5,0) circle(0.1);
				\draw[fill=white] (2,0) circle(0.1);
				\draw[fill=black] (0.5,0.5) circle(0.1);
			\end{tikzpicture}
			&
			\begin{tikzpicture}
				\draw (0,0)--(0,0.5);
				\draw (0.5,0)--(1,0);
				\draw[fill=white] (0,0) circle(0.1);
				\draw[fill=black] (0.5,0) circle(0.1);
				\draw[fill=white] (1,0) circle(0.1);
				\draw[fill=white] (1.5,0) circle(0.1);
				\draw[fill=white] (2,0) circle(0.1);
				\draw[fill=black] (0,0.5) circle(0.1);
			\end{tikzpicture}
			&
			\begin{tikzpicture}
				\draw (0,0)--(2,0);
				\draw (1.5,0)--(2,0);
				\draw (0.5,0)--(1,0.5);
				\draw (1,0)--(1,0.5);
				\draw[fill=white] (0,0) circle(0.1);
				\draw[fill=white] (0.5,0) circle(0.1);
				\draw[fill=black] (1,0) circle(0.1);
				\draw[fill=white] (1.5,0) circle(0.1);
				\draw[fill=white] (2,0) circle(0.1);
				\draw[fill=black] plot [only marks, mark = square*] coordinates {(1,0.5)};
			\end{tikzpicture}
			&
			\begin{tikzpicture}
				\draw (0,0)--(1,0);
				\draw (1.5,0)--(2,0);
				\draw (0.5,0)--(0.5,0.5);
				\draw[fill=white] (0,0) circle(0.1);
				\draw[fill=white] (0.5,0) circle(0.1);
				\draw[fill=white] (1,0) circle(0.1);
				\draw[fill=white] (1.5,0) circle(0.1);
				\draw[fill=white] (2,0) circle(0.1);
				\draw[fill=black] plot [only marks, mark = square*] coordinates {(0.5,0.5)};
			\end{tikzpicture}
			&
			\begin{tikzpicture}
				\draw (0,0)--(2,0);
				\draw (0.95,0)--(0.95,0.5);
				\draw (1.05,0)--(1.05,0.5);
				\draw[fill=white] (0,0) circle(0.1);
				\draw[fill=white] (0.5,0) circle(0.1);
				\draw[fill=black] (1,0) circle(0.1);
				\draw[fill=white] (1.5,0) circle(0.1);
				\draw[fill=white] (2,0) circle(0.1);
				\draw[fill=black] (1,0.5) circle(0.1);
			\end{tikzpicture}
			&
			\begin{tikzpicture}
				\draw (0,0)--(1,0);
				\draw (1.5,0.05)--(2,0.05);
				\draw (1.5,-0.05)--(2,-0.05);
				\draw (0.5,0)--(1,0.5);
				\draw (1,0)--(1,0.5);
				\draw[fill=white] (0,0) circle(0.1);
				\draw[fill=white] (0.5,0) circle(0.1);
				\draw[fill=black] (1,0) circle(0.1);
				\draw[fill=black] (1.5,0) circle(0.1);
				\draw[fill=black] (2,0) circle(0.1);
				\draw[fill=black] (1,0.5) circle(0.1);
			\end{tikzpicture}
			\\\hline
		\end{tabular}
		\caption{(Generalized) Dynkin diagrams representing the Gram matrices of the $T_0$ lattices for the $\K3$ surfaces $S_1$ (denoted by $[a\,b\,c]$) appearing in solutions with $\Nf \leq 30$. Asterisks $*T_S*$ denote $N_\text{flux}\geq 25$ (cf. Table \ref{tab:rankW2}). We use custom conventions when there are vectors with norm larger than 2: (i) vectors with norm 4 and 6 are respectively represented by black-filled circles and squares, and (ii) the number of links always denotes the inner product of two vectors, i.e. one line corresponds to $ -1$ and two lines to $-2$. For $\footnotesize\begin{bmatrix}4&0&4\\\end{bmatrix}$ the crossing line denotes a positive inner product $+1$.}
		\label{tab:Dynkins}
	\end{center}
\end{table}

\subsection{Finding and examining embeddings of \texorpdfstring{$T_0$}{T0} into\texorpdfstring{$N_I$}{NI}}
Continuing with the Kneser-Nishiyama method, we now have to embed the $T_0$'s in Table \ref{tab:Dynkins} into all the possible Niemeier lattices $N_I$. A necessary condition for this embedding to exist is that the root sublattice $(T_0)_\text{root}$ of $T_0$ is a sublattice of the root sublattice $(N_I)_\text{root}$ of $N_I$. It is clear that this condition can only be satisfied for the Leech lattice $N_\omega$ if $(T_0)_\text{root} = \emptyset$, which is not true for \textit{any} $T_0$ we have obtained such that the tadpole bound is satisfied.

\subsubsection{Algorithm 1: minimum non-Abelian gauge group rank}
\label{sss:alg1}
In general, there are many inequivalent primitive embeddings of $T_0$ into any of the $N_I$, all of which in principle we have to consider in order to obtain the possible frame lattices $W$. These embeddings and their associated $W$'s can be obtained as follows:
\begin{enumerate}
	\item We construct all possible embeddings of $(T_0)_\text{root}$ into $(N_I)_\text{root}$ such that the generators of the former correspond to a subset of the generators of the latter, where we take as generators a set of simple roots for the ADE root systems. In other words, we take the Dynkin diagram of $(T_0)_\text{root}$ to be embedded into the Dynkin diagram of $(N_I)_\text{root}$ (up to automorphisms of $N_I$). By virtue of Propositions \ref{prop:prim} and \ref{prop:roots} of Appendix \ref{app:embd_in_Niemeier}, this procedure is exhaustive, i.e. it gives every possible embedding of $(T_0)_\text{root}$ into $N_I$, all of which are guaranteed to be primitive.
	
	\item It is always the case that $T_0$ is obtained by extending the generating set $\{v_i\}$ of $(T_0)_\text{root}$ by $r-6$ vectors with norm $> 2$, with $r$ the rank of $(T_0)_\text{root}$. In particular,  $(T_0)_\text{root} = T_0$ for $r = 6$. For $r \leq 5$, the extra generators are chosen from a previously generated list containing all the vectors in $N_I$ with the desired norms, such that their products with $\{v_i\}$ correspond to the Gram matrix of $T_0$. Primitivity of the resulting embedding is then checked. 
	\item  For every embedding of $T_0$ obtained we compute its orthogonal complement $W$ in $N_I$. This is done in the same way as how $T_0$ is obtained as the orthogonal complement of $T_S$ in $E_8$, as described above.  
\end{enumerate}

In practice we are not interested in recording the exact data defining every possible frame lattice $W$. What we care about is ``how much" non-Abelian gauge symmetry is realized for each fibration of a given $\K3$, which can be quantified e.g. as the rank of $W_\text{root}$; in particular we care about the overall minimum of this value. For each embedding of $T_0$ into $N_I$ in the above algorithm we may restrict therefore to a computation of this rank, which can be done by finding any maximal set of linearly independent roots of $N_I$ orthogonal to $T_0$ --- its number of elements is just the rank of $W_\text{root}$. 

We have carried out this scan exhaustively for every $T_0$ in Table \ref{tab:Dynkins} appearing in solutions with $N_\text{flux}\leq24$, except for those corresponding to $T_S = [7~7~7]\,,~[8~8~8]$, for which our results are partial at this stage and yield upper bounds on the minimal rank of root sublattices (see below for an alternative treatment). The reason for this is that for larger norms, the number of vectors in a given lattice (in this case the $N_I$) generically increases, and the number of pairs of such vectors even more so. This presents a purely technical problem, as computational times and memory requirements get out of control. The cases we have examined exhaustively using the method just outlined have at most two vectors with norm 4 or one with norm 6. The  values of the minimal rank of the associated $W_\text{root}$'s are presented in Table \ref{tab:rankW2}, together with the aforementioned upper bounds for the exceptional cases.

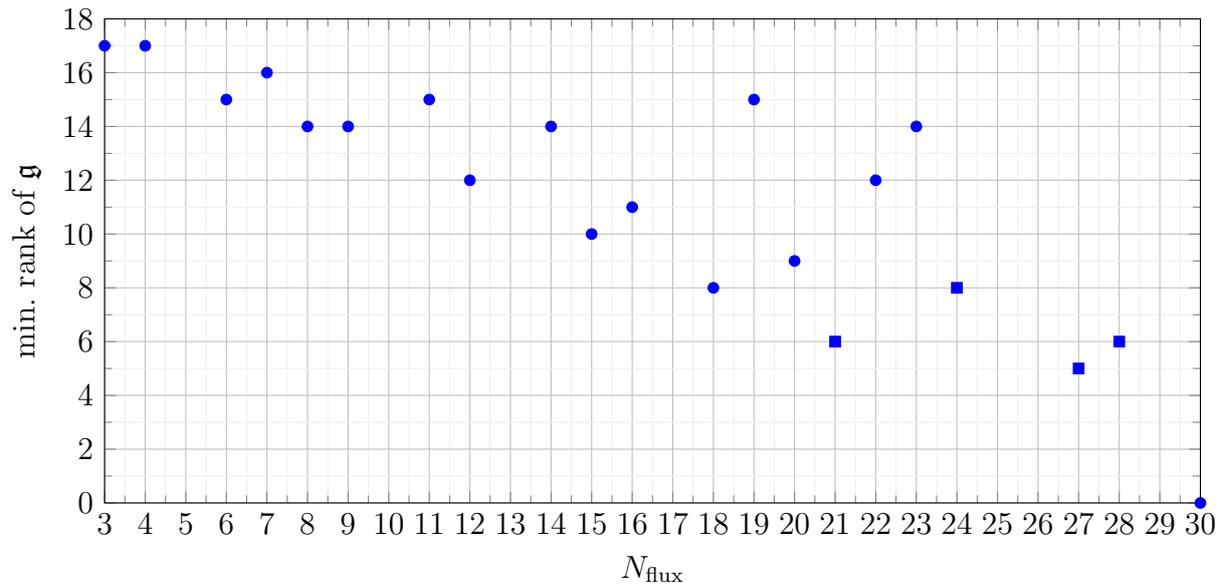
\begin{figure}[H]
	\begin{tikzpicture}
	\begin{axis}[
	xmin = 3, xmax = 30,
	ymin = 0, ymax = 18,
	xtick distance = 1,
	ytick distance = 2,
	grid = both,
	minor tick num = 1,
	major grid style = {lightgray},
	minor grid style = {lightgray!25},
	width = \textwidth,
	height = 0.5\textwidth,
	xlabel = {$N_\text{flux}$},
	ylabel = {$\text{min. rank of }\mathfrak{g}$},]
	
	\addplot[ blue, only marks] table [x = {x}, y = {y}] {
		x y
		3 17
		4 17
		6 15
		7 16
		8 14
		9 14
		11 15
		12 12
		14 14
		15 10
		16 11
		18 8
		19 15
		20 9
		22 12
		23 14
		30 0
	};
	\addplot[ blue, only marks, mark = square*] table [x = {x}, y = {y}] {
		x y
		21 6
		24 8
		27 5
		28 6
	};
	\end{axis}
	\end{tikzpicture}
	\caption{Values of the minimum rank of non-Abelian gauge algebras $\mathfrak{g}$ realizable by backgrounds with given $N_\text{flux}$. The squares represent upper bounds (cf. Table \ref{tab:rankW2}). For all cases except at $N_\text{flux} = 30$ the rank of $\mathfrak{g}$ is strictly larger than 0 (see also the discussion in Section~\ref{sss:alg2}).}
	\label{fig:rankvflux}
\end{figure}

Having these results at hand we may have expected from the tadpole conjecture that as $N_\text{flux}$ increases, the minimal rank of allowed gauge algebras should decrease. We have plotted this relation in Figure \ref{fig:rankvflux}, which shows that such a trend is present in a very roughly linear fashion (with large deviations at certain points whose $\Nf$ admits only one or two solutions). On the other hand, we do observe a much sharper decrease of this rank as the \textit{determinant} of $T_S$ increases. This data is plotted in Figure \ref{fig:rankvdet}.

We have also explored the non-Abelian gauge symmetries associated to the $T_S$'s considered here and find it interesting that every possible simple ADE gauge algebra $\mathfrak g$ is realized up to the overall constraint $\text{rank}(\mathfrak g) \leq 18$. One could have expected some nontrivial restriction on the algebras which satisfy the tadpole bound; if there is one, it is more subtle than simply disallowing certain ADE types. In Appendix \ref{app:alg} we record samples of possible gauge algebras for each elliptic $\K3$ under consideration.

\subsubsection{Algorithm 2: existence of purely Abelian gauge group}
\label{sss:alg2}
To deal with the $T_0$'s with two vectors of norm 6 we have employed an alternative algorithm which determines only whether or not there exist corresponding frame lattices without roots. Although it is more limited than the above algorithm, its scope of applicability is greater and still answers the main question posed by the tadpole conjecture in this scenario. 

This alternative algorithm takes as an input one of the embeddings of $(T_0)_\text{root}$ constructed in step 1 of the previous algorithm and proceeds as follows:
\begin{enumerate}
	\item Construct a list $\{\alpha_1,...,\alpha_n\}$ with every positive root (using $\alpha_i$ and $-\alpha_i$ is redundant in the following) in $N_I$ orthogonal to $(T_0)_\text{root}$ and two lists $V$ and $W$ each with every norm 6 vector in $N_I$ such that every pair $(v,w)$ satisfying the condition to extend $(T_0)_\text{root}$ to $T_0$ is in $V\times W$.
	
	\item Separate $V$ into two lists $V^\perp_1$ and $V^\|_1$ according to whether the elements of $V$ are orthogonal or not to $\alpha_1$. Similarly, separate $W$ into $W_1^\perp$ and $W_1^\|$. 
	
	\item Construct three tuples $(V^\perp_1,W_1^\|)$, $(V^\|_1,W_1^\perp)$ and $(V^\|_1,W_1^\|)$, which are the only ones containing pairs of norm 6 vectors defining embeddings of $T_0$ not orthogonal to $\alpha_1$. In other words, we discard $(V^\perp_1,W_1^\perp)$ as it automatically leads to frame lattices with at least two roots $\pm \alpha_1$.
	
	\item Perform steps 2 and 3 separately on these three tuples using orthogonality with $\alpha_2$. Iterate until every tuple of the type $(V^\perp,W^\perp)$ has been discarded. If at some iteration there remain no inputs for the next, there exists no frame lattice without roots. Otherwise the algorithm must continue up to the $n$-th iteration, producing one or more tuples. It is still possible that the pairs of norm 6 vectors in such tuples do not properly extend $(T_0)_\text{root}$ to $T_0$, in which case the result is again negative. Otherwise we do obtain explicitly one or more frame lattices without roots.
\end{enumerate} 
This algorithm must be applied for every embedding of $(T_0)_\text{root}$ into every possible $N_I$. If for one of these one finds some $W$ without roots, the algorithm stops. 

Performing this algorithm for $T_S = [7~7~7]\,,~[8~8~8]$ we see that for every embedding of $(T_0)_\text{root}$ it stops at some iteration, thus we conclude that there exists no corresponding $W$ without roots and the tadpole conjecture is verified. 

\subsection{Beyond the tadpole bound}

Having seen that there exists no frame lattice without roots associated to a $T_S$ admitting $N_\text{flux} \leq 24$, it is natural to ask what is the minimum value of $N_\text{flux}$ for which one does exist. 

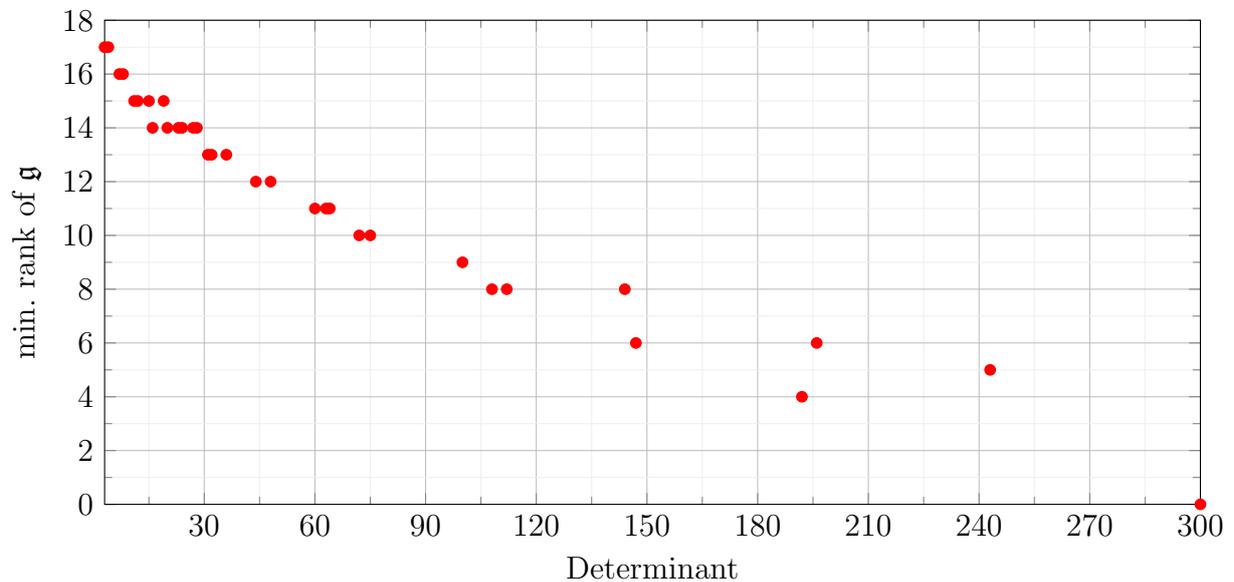
\begin{figure}[H]
	\begin{tikzpicture}
	\begin{axis}[
	xmin = 3, xmax = 300,
	ymin = 0, ymax = 18,
	xtick distance = 30,
	ytick distance = 2,
	grid = both,
	minor tick num = 1,
	major grid style = {lightgray},
	minor grid style = {lightgray!25},
	width = \textwidth,
	height = 0.5\textwidth,
	xlabel = {Determinant},
	ylabel = {$\text{min. rank of }\mathfrak{g}$},]
	
	\addplot[ red, only marks] table [x = {x}, y = {y}] {
		x y
		3 17
		4 17
		7 16
		8 16
		11 15
		12 15
		15 15
		16 14
		19 15
		20 14
		23 14
		24 14
		27 14
		28 14
		31 13
		32 13
		36 13
		44 12
		48 12
		60 11
		63 11
		64 11
		72 10
		75 10
		100 9
		108 8 
		112 8
		144 8
		147 6
		192 4
		196 6
		243 5
		300 0
	};
	\end{axis}
	\end{tikzpicture}
	\caption{Values of the minimum rank of non-Abelian gauge algebras $\mathfrak{g}$ associated to elliptic $\K3$s ordered by the determinant of their transcendental lattice $T_S$. }
	\label{fig:rankvdet}
\end{figure}

To this end we consider the $T_S$'s with the next three allowed values $N_\text{flux} = 27,28,30$; their corresponding $T_0$'s are the last eight shown in Table \ref{tab:Dynkins}. Four of these $T_0$'s, corresponding to $T_S = [7~1~1]\,,~[7~0~1]\,,~ [7~0~7]\,,~[4~2~4]$, have the simple form that is amenable to the algorithm of section \ref{sss:alg1}, allowing to find the minimum ranks of the root systems associated to the frame lattices. These are respectively 14, 14, 6 and 11. The minimum values of $N_\text{flux}$ taken by the corresponding backgrounds are 27, 28, 30 and 28. However, these values of $\Nf$ are also obtained using the four remaining $T_S$'s, and so these could push down the minimum ranks as functions of $\Nf$. Indeed these other lattices have larger determinant and from the trend in Figure \ref{fig:rankvdet} we do expect this outcome. 

We have subjected the lattices $T_S = [9~9~9]\,,~[8~4~4]$ to the algorithm of Section \ref{sss:alg2} and found that for these there are no frame lattices without roots, similarly to the cases $[7~7~7]$ and $[8~8~8]$ studied previously. This settles the question of existance of such lattices for $\Nf \leq 28$ giving a negative answer; the relationship between symmetry enhancements and the tadpole bound is not as fine tuned as we could have expected a priori. Instead it turns out that a frame lattice without roots does exist for $T_S = [10~10~10]$, which we construct explicitly in the following.

Consider the Leech lattice $N_\omega$ generated by the row vectors 
\begin{equation} 
	\begin{pmatrix}
		v_1 \\ \vdots \\ v_{24}
	\end{pmatrix} = 
	\left(\begin{smallmatrix}
		~\,\,8 & ~\,\,4 & ~\,\,4 & ~\,\,4 & ~\,\,4 & ~\,\,4 & ~\,\,4 &~\,\,2 & ~\,\,4 & ~\,\,4 & ~\,\,4 &~\,\,2 & ~\,\,4 &~\,\,2 &~\,\,2 &~\,\,2 & ~\,\,4 &~\,\,2 &~\,\,2 &~\,\,2 &~\,\,0 &~\,\,0 &~\,\,0 & -3 \\
		~\,\,4 & ~\,\,4 &~\,\,2 &~\,\,2 &~\,\,2 &~\,\,2 &~\,\,2 &~\,\,2 &~\,\,2 &~\,\,2 &~\,\,2 &~\,\,2 &~\,\,2 &~\,\,2 &~\,\,1 &~\,\,1 &~\,\,2 &~\,\,1 &~\,\,1 &~\,\,2 &~\,\,1 &~\,\,0 &~\,\,0 & -1 \\
		~\,\,4 &~\,\,2 & ~\,\,4 &~\,\,2 &~\,\,2 &~\,\,2 &~\,\,2 &~\,\,2 &~\,\,2 &~\,\,2 &~\,\,2 &~\,\,2 &~\,\,2 &~\,\,1 &~\,\,2 &~\,\,1 &~\,\,2 &~\,\,2 &~\,\,1 &~\,\,1 &~\,\,1 &~\,\,0 &~\,\,0 & -1 \\
		~\,\,4 &~\,\,2 &~\,\,2 & ~\,\,4 &~\,\,2 &~\,\,2 &~\,\,2 &~\,\,2 &~\,\,2 &~\,\,2 &~\,\,2 &~\,\,2 &~\,\,2 &~\,\,1 &~\,\,1 &~\,\,2 &~\,\,2 &~\,\,1 &~\,\,2 &~\,\,1 &~\,\,1 &~\,\,0 &~\,\,0 & -1 \\
		~\,\,4 &~\,\,2 &~\,\,2 &~\,\,2 & ~\,\,4 &~\,\,2 &~\,\,2 &~\,\,2 &~\,\,2 &~\,\,2 &~\,\,2 &~\,\,1 &~\,\,2 &~\,\,2 &~\,\,2 &~\,\,2 &~\,\,2 &~\,\,2 &~\,\,2 &~\,\,2 &~\,\,1 &~\,\,0 &~\,\,0 & -1 \\
		~\,\,4 &~\,\,2 &~\,\,2 &~\,\,2 &~\,\,2 & ~\,\,4 &~\,\,2 &~\,\,2 &~\,\,2 &~\,\,2 &~\,\,2 &~\,\,1 &~\,\,2 &~\,\,2 &~\,\,1 &~\,\,1 &~\,\,2 &~\,\,1 &~\,\,2 &~\,\,1 &~\,\,0 &~\,\,0 &~\,\,0 & -1 \\
		~\,\,4 &~\,\,2 &~\,\,2 &~\,\,2 &~\,\,2 &~\,\,2 & ~\,\,4 &~\,\,2 &~\,\,2 &~\,\,2 &~\,\,2 &~\,\,1 &~\,\,2 &~\,\,1 &~\,\,2 &~\,\,1 &~\,\,2 &~\,\,1 &~\,\,1 &~\,\,2 &~\,\,0 &~\,\,0 &~\,\,0 & -1 \\
		~\,\,2 &~\,\,2 &~\,\,2 &~\,\,2 &~\,\,2 &~\,\,2 &~\,\,2 & ~\,\,4 &~\,\,1 &~\,\,1 &~\,\,1 &~\,\,2 &~\,\,1 &~\,\,2 &~\,\,2 &~\,\,2 &~\,\,1 &~\,\,2 &~\,\,2 &~\,\,2 &~\,\,2 &~\,\,0 &~\,\,0 & ~\,\,1 \\
		~\,\,4 &~\,\,2 &~\,\,2 &~\,\,2 &~\,\,2 &~\,\,2 &~\,\,2 &~\,\,1 & ~\,\,4 &~\,\,2 &~\,\,2 &~\,\,2 &~\,\,2 &~\,\,2 &~\,\,2 &~\,\,2 &~\,\,2 &~\,\,2 &~\,\,2 &~\,\,2 &~\,\,1 &~\,\,1 &~\,\,1 & -1 \\
		~\,\,4 &~\,\,2 &~\,\,2 &~\,\,2 &~\,\,2 &~\,\,2 &~\,\,2 &~\,\,1 &~\,\,2 & ~\,\,4 &~\,\,2 &~\,\,2 &~\,\,2 &~\,\,2 &~\,\,1 &~\,\,1 &~\,\,2 &~\,\,2 &~\,\,1 &~\,\,1 &~\,\,0 &~\,\,1 &~\,\,0 & -1 \\
		~\,\,4 &~\,\,2 &~\,\,2 &~\,\,2 &~\,\,2 &~\,\,2 &~\,\,2 &~\,\,1 &~\,\,2 &~\,\,2 & ~\,\,4 &~\,\,2 &~\,\,2 &~\,\,1 &~\,\,2 &~\,\,1 &~\,\,2 &~\,\,1 &~\,\,2 &~\,\,1 &~\,\,0 &~\,\,0 &~\,\,1 & -1 \\
		~\,\,2 &~\,\,2 &~\,\,2 &~\,\,2 &~\,\,1 &~\,\,1 &~\,\,1 &~\,\,2 &~\,\,2 &~\,\,2 &~\,\,2 & ~\,\,4 &~\,\,1 &~\,\,2 &~\,\,2 &~\,\,2 &~\,\,1 &~\,\,2 &~\,\,2 &~\,\,2 &~\,\,2 &~\,\,1 &~\,\,1 & ~\,\,1 \\
		~\,\,4 &~\,\,2 &~\,\,2 &~\,\,2 &~\,\,2 &~\,\,2 &~\,\,2 &~\,\,1 &~\,\,2 &~\,\,2 &~\,\,2 &~\,\,1 & ~\,\,4 &~\,\,2 &~\,\,2 &~\,\,2 &~\,\,2 &~\,\,1 &~\,\,1 &~\,\,1 &~\,\,1 &~\,\,1 &~\,\,1 & -1 \\
		~\,\,2 &~\,\,2 &~\,\,1 &~\,\,1 &~\,\,2 &~\,\,2 &~\,\,1 &~\,\,2 &~\,\,2 &~\,\,2 &~\,\,1 &~\,\,2 &~\,\,2 & ~\,\,4 &~\,\,2 &~\,\,2 &~\,\,1 &~\,\,2 &~\,\,2 &~\,\,2 &~\,\,2 &~\,\,2 &~\,\,1 & ~\,\,1 \\
		~\,\,2 &~\,\,1 &~\,\,2 &~\,\,1 &~\,\,2 & ~\,\,1 &~\,\,2 &~\,\,2 &~\,\,2 &~\,\,1 &~\,\,2 &~\,\,2 &~\,\,2 &~\,\,2 & ~\,\,4 &~\,\,2 &~\,\,1 &~\,\,2 &~\,\,2 &~\,\,2 &~\,\,2 &~\,\,1 &~\,\,2 & ~\,\,1 \\
		~\,\,2 &~\,\,1 &~\,\,1 &~\,\,2 &~\,\,2 &~\,\,1 &~\,\,1 &~\,\,2 &~\,\,2 &~\,\,1 &~\,\,1 &~\,\,2 &~\,\,2 &~\,\,2 &~\,\,2 & ~\,\,4 &~\,\,1 &~\,\,2 &~\,\,2 &~\,\,2 &~\,\,2 &~\,\,1 &~\,\,1 & ~\,\,1 \\
		~\,\,4 &~\,\,2 &~\,\,2 &~\,\,2 &~\,\,2 &~\,\,2 &~\,\,2 &~\,\,1 &~\,\,2 &~\,\,2 &~\,\,2 &~\,\,1 &~\,\,2 &~\,\,1 &~\,\,1 &~\,\,1 & ~\,\,4 &~\,\,2 &~\,\,2 &~\,\,2 &~\,\,1 &~\,\,1 &~\,\,1 & -1 \\
		~\,\,2 &~\,\,1 &~\,\,2 &~\,\,1 &~\,\,2 &~\,\,1 &~\,\,1 &~\,\,2 &~\,\,2 &~\,\,2 &~\,\,1 &~\,\,2 &~\,\,1 &~\,\,2 &~\,\,2 &~\,\,2 &~\,\,2 & ~\,\,4 &~\,\,2 &~\,\,2 &~\,\,2 &~\,\,2 &~\,\,1 & ~\,\,1 \\
		~\,\,2 &~\,\,1 &~\,\,1 &~\,\,2 &~\,\,2 &~\,\,2 &~\,\,1 &~\,\,2 &~\,\,2 &~\,\,1 &~\,\,2 &~\,\,2 &~\,\,1 &~\,\,2 &~\,\,2 &~\,\,2 &~\,\,2 &~\,\,2 & ~\,\,4 &~\,\,2 &~\,\,2 &~\,\,1 &~\,\,2 & ~\,\,1 \\
		~\,\,2 &~\,\,2 &~\,\,1 &~\,\,1 &~\,\,2 &~\,\,1 &~\,\,2 &~\,\,2 &~\,\,2 &~\,\,1 &~\,\,1 &~\,\,2 &~\,\,1 &~\,\,2 &~\,\,2 &~\,\,2 &~\,\,2 &~\,\,2 &~\,\,2 & ~\,\,4 &~\,\,2 &~\,\,1 &~\,\,1 & ~\,\,1 \\
		~\,\,0 &~\,\,1 &~\,\,1 &~\,\,1 &~\,\,1 &~\,\,0 &~\,\,0 &~\,\,2 &~\,\,1 &~\,\,0 &~\,\,0 &~\,\,2 &~\,\,1 &~\,\,2 &~\,\,2 &~\,\,2 &~\,\,1 &~\,\,2 &~\,\,2 &~\,\,2 & ~\,\,4 &~\,\,2 &~\,\,2 & ~\,\,2 \\
		~\,\,0 &~\,\,0 &~\,\,0 &~\,\,0 &~\,\,0 &~\,\,0 &~\,\,0 &~\,\,0 &~\,\,1 &~\,\,1 &~\,\,0 &~\,\,1 &~\,\,1 &~\,\,2 &~\,\,1 &~\,\,1 &~\,\,1 &~\,\,2 &~\,\,1 &~\,\,1 &~\,\,2 & ~\,\,4 &~\,\,2 & ~\,\,2 \\
		~\,\,0 &~\,\,0 &~\,\,0 &~\,\,0 &~\,\,0 &~\,\,0 &~\,\,0 &~\,\,0 &~\,\,1 &~\,\,0 &~\,\,1 &~\,\,1 &~\,\,1 &~\,\,1 &~\,\,2 &~\,\,1 &~\,\,1 &~\,\,1 &~\,\,2 &~\,\,1 &~\,\,2 &~\,\,2 & ~\,\,4 & ~\,\,2 \\
		-3 & -1 & -1 & -1 & -1 & -1 & -1 & ~\,\,1 & -1 & -1 & -1 & ~\,\,1 & -1 & ~\,\,1 & ~\,\,1 & ~\,\,1 & -1 & ~\,\,1 & ~\,\,1 & ~\,\,1 & ~\,\,2 & ~\,\,2 & ~\,\,2 & ~\,\,4 \\
	\end{smallmatrix}\right)\,.
\end{equation}
It can be checked that the vectors
\begin{equation}
	v_{15}-v_7,~~~v_1-v_2,~~~ v_{12}-v_{16}, ~~~ v_1-v_3,~~~v_1-v_5, ~~~ v_{11}
\end{equation}
generate a sublattice
\begin{equation}
	T_0 = \begin{pmatrix}
		~\,\,4 & -1 & ~\,\,0 & -2 & -2 & ~\,\,0 \\
		-1 & ~\,\,4 & -1 & ~\,\,2 & ~\,\,2 & ~\,\,2 \\
		~\,\,0 & -1 & ~\,\,4 & -1 & ~\,\,1 & ~\,\,1 \\
		-2 & ~\,\,2 & -1 & ~\,\,4 & ~\,\,2 & ~\,\,2 \\
		-2 & ~\,\,2 & ~\,\,1 & ~\,\,2 & ~\,\,4 & ~\,\,2 \\
		~\,\,0 & ~\,\,2 & ~\,\,1 & ~\,\,2 & ~\,\,2 & ~\,\,4 \\
	\end{pmatrix}\,.
\end{equation}
This $T_0$ is complementary to $T_S = [10~10~10]$, and since it is embedded into the Leech lattice its orthogonal complement automatically has no  roots. We also note that there is an alternative $T_0$ which takes a simpler form, recorded in Table \ref{tab:Dynkins} in the last entry. It can be seen that it admits an embedding into the Niemeier $N_\chi$ with $(N_\chi)_\text{root} = 12\, A_2$ such that its orthogonal complement has no roots.

Having shown that for $\Nf = 30$ there exists a frame lattice without roots, it becomes unnecesary to study the remaining lattice $T_S = [8~2~2]$. Our analysis is then complete; we see that, although not exactly above the tadpole bound, not too far from it there do exist backgrounds without non-Abelian gauge symmetries.

\begin{table}[htbp]
	\footnotesize
	\begin{center}
		\begin{tabular}{||>{$}c<{$}|>{$}c<{$}|>{$}c<{$}|>{$}c<{$}||}
			\hline
		T_S & \text{Det} & 
		\text{Min } \text{rank}
		& \text{Min tadpole} \\ \hline
		\hline	\begin{array}{ccc}
		1 & 1 & 1 \\
	\end{array}
	& 3 & 17 & 3 \\
	
\hline 	\begin{array}{ccc}
		1 & 0 & 1 \\
	\end{array}
	& 4 & 17 & 4 \\
	
\hline 	\begin{array}{ccc}
		2 & 1 & 1 \\
	\end{array}
	& 7 & 16 & 7 \\
	
\hline 	\begin{array}{ccc}
		2 & 0 & 1 \\
	\end{array}
	& 8 & 16 & 8 \\
	
\hline 	\begin{array}{ccc}
		3 & 1 & 1 \\
	\end{array}
	& 11 & 15 & 11 \\
	
\hline 	\begin{array}{ccc}
		2 & 2 & 2 \\
	\end{array}
	& 12 & 15 & 6 \\
	
\hline 	\begin{array}{ccc}
		3 & 0 & 1 \\
	\end{array}
	& 12 & 15 & 12 \\
	
\hline 	\begin{array}{ccc}
		2 & 1 & 2 \\
	\end{array}
	& 15 & 15 & 15 \\
	
\hline 	\begin{array}{ccc}
		4 & 1 & 1 \\
	\end{array}
	& 15 & 15 & 15 \\
	
\hline 	\begin{array}{ccc}
		2 & 0 & 2 \\
	\end{array}
	& 16 & 14 & 8 \\
	
\hline 	\begin{array}{ccc}
		4 & 0 & 1 \\
	\end{array}
	& 16 & 15 & 16 \\
	
\hline 	\begin{array}{ccc}
		5 & 1 & 1 \\
	\end{array}
	& 19 & 15 & 19 \\
	
\hline 	\begin{array}{ccc}
		3 & 2 & 2 \\
	\end{array}
	& 20 & 14 & 20 \\
	
\hline 	\begin{array}{ccc}
		5 & 0 & 1 \\
	\end{array}
	& 20 & 14 & 20 \\
	
\hline 	\begin{array}{ccc}
		3 & 1 & 2 \\
	\end{array}
	& 23 & 14 & 23 \\
	
\hline 	\begin{array}{ccc}
		6 & 1 & 1 \\
	\end{array}
	& 23 & 14 & 23 \\
	
\hline 	\begin{array}{ccc}
		3 & 0 & 2 \\
	\end{array}
	& 24 & 14 & 24 \\
	
\hline 	\begin{array}{ccc}
		6 & 0 & 1 \\
	\end{array}
	& 24 & 14 & 24 \\
	
\hline 	\begin{array}{ccc}
		3 & 3 & 3 \\
	\end{array}
	& 27 & 14 & 9 \\
	
\hline 	\begin{array}{ccc}
		7 & 1 & 1 \\
	\end{array}
	& 27 & 14 & 27 \\
	
\hline 	\begin{array}{ccc}
		4 & 2 & 2 \\
	\end{array}
	& 28 & 14 & 14 \\
	
\hline 
\end{tabular}
		\begin{tabular}{||>{$}c<{$}|>{$}c<{$}|>{$}c<{$}|>{$}c<{$}||}
			\hline
			T_S & \text{Det} &
\text{Min } \text{rank}
			  & \text{Min tadpole} \\ \hline\hline
			  \begin{array}{ccc}
			  	7 & 0 & 1 \\
			  \end{array}
			  & 28 & 14 & 28 \\
		\hline 	\begin{array}{ccc}
			4 & 0 & 2 \\
		\end{array}
		& 32 & 13 & 16 \\
		
		\hline 	\begin{array}{ccc}
			3 & 0 & 3 \\
		\end{array}
		& 36 & 13 & 12 \\
		
		\hline 	\begin{array}{ccc}
			6 & 2 & 2 \\
		\end{array}
		& 44 & 12 & 22 \\
		
		\hline 	\begin{array}{ccc}
			4 & 4 & 4 \\
		\end{array}
		& 48 & 12 & 12 \\
		
		\hline 	\begin{array}{ccc}
			6 & 0 & 2 \\
		\end{array}
		& 48 & 12 & 24 \\
		
		\hline 	\begin{array}{ccc}
			4 & 2 & 4 \\
		\end{array}
		& 60 & \text{11} & 30 \\
		
		\hline 	\begin{array}{ccc}
			8 & 2 & 2 \\
		\end{array}
		& 60 & \text{$\leq $13} & 30 \\
		
		\hline 	\begin{array}{ccc}
			6 & 3 & 3 \\
		\end{array}
		& 63 & 11 & 21 \\
		
		\hline 	\begin{array}{ccc}
			4 & 0 & 4 \\
		\end{array}
		& 64 & 11 & 16 \\
		
		\hline 	\begin{array}{ccc}
			6 & 0 & 3 \\
		\end{array}
		& 72 & 10 & 24 \\
		
		\hline 	\begin{array}{ccc}
			5 & 5 & 5 \\
		\end{array}
		& 75 & 10 & 15 \\
		
		\hline 	\begin{array}{ccc}
			5 & 0 & 5 \\
		\end{array}
		& 100 & 9 & 20 \\
		
		\hline 	\begin{array}{ccc}
			6 & 6 & 6 \\
		\end{array}
		& 108 & \text{8} & 18 \\
		
		\hline 	\begin{array}{ccc}
			8 & 4 & 4 \\
		\end{array}
		& 112 & \text{$1 \leq r \leq $10} & 28 \\
		
		\hline 	\begin{array}{ccc}
			6 & 0 & 6 \\
		\end{array}
		& 144 & 8 & 24 \\
		
		\hline 	\begin{array}{ccc}
			7 & 7 & 7 \\
		\end{array}
		& 147 & \text{1 $\leq r \leq $6} & 21 \\
		
		\hline 	\begin{array}{ccc}
			8 & 8 & 8 \\
		\end{array}
		& 192 & \text{1 $\leq r \leq $4} & 24 \\
		
		\hline 	\begin{array}{ccc}
			7 & 0 & 7 \\
		\end{array}
		& 196 & \text{6} & 28 \\
		
		\hline 	\begin{array}{ccc}
			9 & 9 & 9 \\
		\end{array}
		& 243 & \text{1$\leq r \leq $5} & 27 \\
		
		\hline 	\begin{array}{ccc}
			10 & 10 & 10 \\
		\end{array}
		& 300 & 0 & 30 \\ \hline	
\end{tabular}		
\caption{Smallest rank for the $W_{\text{root}}$ for the attractive $\K3$ surfaces with $N_\text{flux}\leq 30$, labeled by $T_S$. We give higher bounds for the cases that we have not explored exhaustively. }\label{tab:rankW2} 
\end{center}
\end{table}

\section{Conclusions}

In this paper we have considered the set of F-Theory compactifications on $\Ktt$ with supersymmetric four-form fluxes of a particular simple form, given in \eqref{eq:fluxpuretype0}.  These lead to attractive $\K3$ surfaces with fully stabilised complex structure moduli. The list of all solutions with fluxes below and slightly above the tadpole bound $\Nf \leq 24$ is given in Table \ref{tab:AK-morethan24}. We have shown explicitly that whenever the tadpole bound is satisfied, these vacua always exhibit non-Abelian gauge symmetries.  This implies that moduli stabilisation within the tadpole bound happens at special points in the moduli space, in line with the tadpole conjecture. Furthermore, in order not to have non-Abelian gauge algebra we need to go all the way to $\Nf=30$.

Our results were obtained by employing the Kneser-Nishiyama method, which allows to find the frame lattice of the elliptic fibration by embedding into Niemeier lattices. By this method we have scanned all the possible frame lattices of the elliptic $\K3$s admitted by fluxes within the tadpole bound, showing that they always contain roots. Once we relax the tadpole bound, one can find F-theory flux solutions without non-abelian gauge groups. As the rank of the gauge group including abelian factors is always $18$ for the solutions we consider, the associated $\K3$ surfaces have a Mordell-Weil lattice of rank $18$.

We saw that the algebras appearing at the vacua with $\Nf \leq 24$ do not seem to be constrained in any evident way; in particular every simple ADE factor up to rank 18 can be found among the whole list. We have recorded some of these algebras in Appendix \ref{app:alg}. It would be interesting to see if a nontrivial constraint on the gauge algebras exists.

We have also shown with an explicit construction that for $\Nf = 30$ there exists a vacuum with no non-Abelian gauge symmetry, in line with the intuition that such vacua should exist not far above the tadpole bound.

One natural extension of this work would involve considering more general fluxes involving the $(1,1)$-forms on each of the $\K3$, namely fluxes of the form  \eqref{genG4}. Such generic fluxes are much harder to analyse, as they are not expected to lead to attractive $\K3$ surfaces.
These flux solutions were actually considered as the original evidence for the tadpole conjecture. Using evolutionary algorithms many {\it M-theory} flux configurations with $\Nf=25$  stabilising all moduli (K\"ahler and complex structure) at generic points in moduli space were constructed, where none were found within the tadpole bound. It would be interesting to see if any of these solutions can be lifted to F-theory.

There is an interesting link between our work and the theory of sphere packings which allows to conjecture this result \cite{Manning-Coe:2023wjl}. For an $n$-dimensional lattice sphere 
packing based on a lattice $L$ with shortest lattice vector of length $v^2 = \lambda$, the centre density of the packing is 
\begin{equation}
 \delta_L = \frac{\lambda^{n/2}}{4} \frac{1}{\sqrt{\det(L)}}\, .
\end{equation}
The tightest known such packing for $n=18$ comes from the lattice $\Lambda_{18}$, which has $\lambda = 4$ and $\det L = 192$ \cite{conway1998sphere,lambda18}. The discriminant group of this lattice is $\Z_2^5 \times \Z_6$, so that it cannot appear as a frame lattice of an attractive $\K3$ surface. This is the tightest sublattice
of the Leech lattice and the tightest known lattice in $18$ dimensions, but no proof of optimality has been given. Note that $192$ is exactly the highest determinant 
among the lattices appearing within the tadpole bound. Hence finding a frame lattice without roots (so that the minimal length vector is $4$) would have given us a new record sphere packing in $18$ dimensions.

\vspace{0.5cm}

\noindent
{\bf Acknowledgments} We would like to thank Iosif Bena and Wati Taylor for useful discussions.
This work was supported in part by the ERC Grant 772408 ``Stringlandscape''. 
The work of SL was in part supported by the NSF grant PHY-1915071.
The work of BF was in part supported by the ERC starting Grant QGuide.

\newpage
\appendix

\section{Lattices}
\label{app:lattices}

In this appendix we state and develop properties of lattices needed for our analysis. References for discriminants and primitive embeddings 
are \cite{MR525944,Morrison1984}. More background on ADE root lattices, their discriminants and embeddings as well as the Niemeier lattices can be found in \cite{alma9944025973902959,conway1998sphere}, see also \cite{nishiyama_1,nishiyama_2} for the present context. 

We will use the term \emph{lattice} $\Lambda$ to refer to a finitely generated free Abelian group together with an integral bilinear 
form $\cdot$, i.e. for all $l,l' \in \Lambda$, $l\cdot l' \in \Z$. Here, \emph{free} means that 
$n l \neq0$ for every $l \neq 0$ and all $n \in \Z$ with $n \neq 0$. This implies that as an Abelian group (i.e. forgetting 
the bilinear form) $\Lambda \cong \Z^r$. The integer $r$ is called the \emph{rank} of $\Lambda$. Choosing a $\Z$-basis $\{l_i\}$ of $\Lambda$ we can write the bilinear 
form as $l_i \cdot l_j = \Omega_{ij}$. The matrix with components $\Omega_{ij}$ is called the \emph{Gram matrix} of the lattice. 
If the rank of the matrix $\Omega$ is $r$, the difference $l_+-l_-$ between the number of positive ($l_+$) and negative eigenvalues ($l_-$) of $\Omega$ is called the \emph{signature} of $\Lambda$. A lattice is called \emph{even} if $l \cdot l \in 2\Z$ for all $l \in \Lambda$ and \emph{odd} otherwise. 

For a lattice $\Lambda$, $\Lambda[k]$ denotes the lattice found by rescaling $\Omega_{ij}$ by $k$.  

Given two lattices $\Lambda$ and $\Psi$ of equal rank for which $\Psi \subset \Lambda$, $\Lambda$ is called an \emph{overlattice} of $\Psi$.

\subsection{Dual Lattice, Discriminant Forms, and Overlattices}

By tensoring with the rationals $\Lambda_\Q := \Lambda \otimes \Q$ becomes a 
vector space, and the bilinear form between lattice elements naturally extends to $\Lambda_\Q$. The \emph{dual lattice} $\Lambda^*$ is the subset of $\Lambda_\Q$ that has an integral product 
with all elements of
$\Lambda$:
\begin{equation}
\Lambda^* = \{ \ell \in \Lambda_\Q  | \ell \cdot l \in \Z \,\,\forall l \in \Lambda\} \, .
\end{equation}
We can use the basis $\{l_i\}$ to express elements of $\Lambda^*$ as well, but then the coefficients will in general 
be rational rather than integer numbers. As $l\cdot l' \in \Z$ for all $l,l' \in \Lambda$, it follows that $\Lambda \subseteq \Lambda^*$. 

As $\Lambda \subseteq \Lambda^*$ we can consider the quotient 
\begin{equation}
G_{\Lambda} :=  \Lambda^*/\Lambda \, ,
\end{equation}
which is called the \emph{discriminant group} of $\Lambda$. We denote the minimal number of generators of $G_{\Lambda}$ by 
$\ell(G_{\Lambda})$. The orders of the generators of $G_{\Lambda}$ are equal to the diagonal entries of the Smith normal form of 
$\Omega_{ij}$ which implies that $\ell(G_{\Lambda}) \leq \rk \Lambda$.

As $\Lambda^*$ is contained in  $\Lambda_\Q$, we can 
extend the bilinear form to $\Lambda^*$ (where it ceases to be integral in general) and hence to $G_{\Lambda}$. 
For $\gamma,\gamma' \in G_{\Lambda}$ we have that 
\begin{equation}
q_\Lambda(\gamma,\gamma') = \gamma \cdot \gamma'\,\ \mbox{mod}\,\, 2\Z  \, ,
\end{equation}
which is called the \emph{discriminant form} of $\Lambda$.

When $\Lambda^* = \Lambda$ the lattice $\Lambda$ is called \emph{self-dual} or \emph{unimodular}. This implies that $\det(\Omega) = \pm 1$. A simple example of an even unimodular lattice is given by the \emph{hyperbolic lattice} $U$ with inner form
\begin{equation} \label{U}
U = \left(\begin{array}{cc}
     0 & 1 \\
     1 & 0
    \end{array}\right) \, .
\end{equation}
This is the unique even unimodular lattice of signature $(1,1)$.

\begin{THM}
An even and self-dual lattice of signature $(p,q)$ exists if and only if $p-q=0\, \mbox{mod}\, 8$. If furthermore both $p \neq 0$
and $q \neq 0$, this lattice is unique (up to isomorphism).  
\end{THM}

For even lattices of definite signature there is a unique self-dual lattice of rank eight, the root lattice $E_8$. For rank $16$ there are two 
such lattices, $E_8 \oplus E_8$ and $\tilde{D}_{16}$, which is an overlattice of $D_{16}$. In dimension $24$, there are $24$ even self-dual lattices, which are 
called the Niemeier lattices. We review their construction in Appendix \ref{app:niemeier}.

\subsection{Primitive Embeddings}\label{app:primitive_embeddings}

For a sublattice $M \subset \Lambda $ the embedding of $M$ is called \emph{primitive} if the quotient
$\Lambda/M$ is free, i.e. is again a lattice. This implies that for every $\ell \in \Lambda$ such that 
$\ell \notin M$, it cannot happen that there is an $n \in \Z$, $ n\neq 0$, such that $n \ell \in M$, as 
this would imply that $\ell \neq 0$, but  $n\ell = 0$ in the quotient. Primitivity of an embedding is equivalent to 
$M \cap \left(\Lambda \otimes \Q\right) = M$. For non-primitive embeddings, the quotient $\Lambda/M$ contains finite groups, which 
are called the torsional subgroup $\tors(\Lambda/M)$. 

\begin{THM}\label{thm:existence_of_emb}
For an even lattice $M$ of signature $(m_+,m_-)$ there exists a primitive embedding into an even self-dual lattice $\Lambda$ of signature $(l_+,l_-)$
if $m_\pm \leq l_\pm$ and $\rk(M) \leq \rk(\Lambda)/2$. 
\end{THM}

\begin{THM}\label{thm:uniqueness_of_emb}
For an even lattice $M$ of signature $(m_+,m_-)$ there exists a unique primitive embedding into an even self-dual lattice $\Lambda$ of signature $(l_+,l_-)$
if $m_\pm < l_\pm$ and $\ell(G_M) \leq \rk(\Lambda) - \rk(M) - 2$. 
\end{THM}

For any embedding, we may consider the \emph{orthogonal complement} 
\begin{equation}
\Lambda^\perp = \{\ell \in M | \ell \cdot l = 0 \,\, \forall \, l \in \Lambda\} \,. 
\end{equation}
The orthogonal complement is automatically primitively embedded in $M$. 

\begin{THM}\label{thm:orthcomp}
In case both $M$ and $M^\perp$ are primitively embedded into an even unimodular lattice $\Lambda$, it follows that 
$G_M \cong G_{M^\perp}$ and
\begin{equation}
q(M) = - q(M^\perp) \, . 
\end{equation} 
The converse is also true: for any pair of even lattices $M$ and $N$ such that $G_M \cong G_N$ with $q(M) = - q(N)$ 
and which furthermore obey $m_++n_+-m_--n_- = 0 \, \, \mbox{mod} \, \, 8$ there exists 
an even unimodular lattice $\Lambda$ such that $M$ and $N$ are primitively embedded into $\Lambda$
and $M = N^\perp$, $N = M^\perp$ in $\Lambda$. 
\end{THM}

For a primitive sublattice $\Psi$ of a lattice $\Lambda$ we always have
\begin{equation}
\Lambda \supseteq \Psi \oplus \Psi^\perp 
\end{equation}
but the above is rarely an equality. An exception to this is when $\Psi$ is self-dual:

\begin{THM}
Let $M$ be a self-dual lattice which is primitively embedded into a lattice $\Lambda$. Then
\beq
\Lambda = M \oplus M^\perp
\eeq  
\end{THM}

\subsection{Root Lattices}\label{app:root_lattices}

For a lattice, we shall call those elements $v$ with $v^2=2$ roots\footnote{These are the conventions natural in group theory. In geometry, 
we will encouter such lattices with a relative minus sign in front of the inner form.}. For any even lattice $\Lambda$, $\Lambda_\text{root} \subset \Lambda$ is the sublattice generated by all roots of $\Lambda$. We can always write
\begin{equation}
\Lambda_\text{root} =  \Gamma_1 \oplus \Gamma_2 \oplus \cdots 
\end{equation}
where $\Gamma_k  \in \{A_n, D_n, E_6,E_7,E_8\}$ is an ADE root lattice, the details of which are described below. For the construction of Niemeier lattices we will need specific elements of the dual lattice `glue vectors') which are also defined below.\\

$\mathbf{A_n}$ \textbf{lattice}:  
\begin{equation}
A_n = \{ v \in \Z^{n+1} | \sum v_i = 0  \}\,.
\end{equation}
Roots $\alpha_{ij}$ are of the form $v_i = 1$, $v_j=-1$ and $0$ else, and the dual lattice $A_n^\ast$ contains $A_n$ together with multiples of 
\begin{equation}
v[1] = \left(\frac{1}{n+1},\cdots,\frac{1}{n+1},-\frac{n}{n+1} \right) \, .
\end{equation}
Clearly $\sum_i (v[1])_i = 0$ and $v[1]  \cdot \alpha_{ij} = 0$ or $1$. As $(n+1) v[1]  \in A_n$ it follows that $G_{A_n} =\Z_{n+1}$ and 
\begin{equation}
q_{A_n} = \frac{n+ n^2}{(n+1)^2} = \frac{n}{n+1}\,.
\end{equation}

The glue vectors we will need are defined as
\begin{equation}
v[i] = \left(\frac{i}{n+1},\cdots,\frac{i}{n+1},-\frac{j}{n+1},\cdots,-\frac{j}{n+1} \right) \, ,
\end{equation}
where $i+j = n+1$, $0 \leq i \leq n$ and the above expression has $j$ components equal to $i/(n+1)$ and $i$ components equal to 
$-j/(n+1)$. 

$\mathbf{D_n}$ \textbf{lattice}:
\begin{equation}
D_n = \{ v \in \Z^{n} | \sum v_i \in 2 \Z  \}\,.
\end{equation}
Roots $\alpha_{ij \pm}$ are of the form $v_i = 1$, $v_j=\pm 1$ and $0$ else. The dual lattice $D_n^\ast$ is generated by the $\alpha_{ij\pm}$ together with 
\begin{equation}
v_d ({\bf a}) = \left((-1)^{a_1} \frac{1}{2},(-1)^{a_2} \frac{1}{2},\cdots (-1)^{a_n} \frac{1}{2}\right) 
\end{equation}
for arbitrary integers $a_i$. These come in two parity types, those with an even number of $-$ signs and those with an odd number of $-$ signs. 
Adding or subtracting appropriate roots of $D_n$ does not change this parity.

If $\mathbf{n}$ is {\bf odd}, $-v_d ({\bf a})$ has the opposite parity to $v_d ({\bf a})$. Modulo $D_n$, there is hence a single generator of $D_n^\ast$. As $n$ is odd
$2v_d ({\bf a}) \notin D_n$, but $4 v_d ({\bf a}) \in D_n$. Hence $G_{A_n} =\Z_{4}$ and
\begin{equation}
q_{D_n} = \frac{n}{4} \, .
\end{equation}

If $\mathbf{n}$ is {\bf even}, we have that $2 v_d ({\bf a}) \in D_n$ for all ${\bf a}$. The different parities of $v_d ({\bf a})$ are no longer equivalent by inverting $v_d ({\bf a})$,
so that $G_{A_n} =\Z_{2} \times \Z_2$ and 
\begin{equation}
q_{D_n} = \begin{pmatrix}
           \frac{n}{4} &  \frac{n-2}{4} \\
              \frac{n-2}{4} & \frac{n}{4}
          \end{pmatrix}\,,
\end{equation}
where we have chosen the generators $v_1 = (\tfrac12^n)$ and $v_2 = (-\tfrac12, \tfrac12^{n-1})$.

The glue vectors we will need are defined as
\begin{equation}
\begin{aligned}
v[1] & = (\tfrac12,\tfrac12,\cdots,\tfrac12)\,, \\
v[2] & = (0,0,\cdots,0,1)\,, \\
v[3] & = (\tfrac12,\tfrac12,\cdots,\tfrac12,-\tfrac12)\,. \\
\end{aligned}
\end{equation}

$\mathbf{E_8}$ \textbf{lattice}:
\begin{equation}
 E_8 = \left\{ v \in \Q^8 \, | \, \sum v_i \in 2 \Z \,\,,\,\,  v_i \in \Z \forall i \,\, \mbox{or}\,\, v_i \in \Z + \tfrac12  \forall i\right\}. 
\end{equation}
It is self-dual, i.e. $E_8^* = E_8$, so that the determinant of the inner form between generators equals one. \\

$\mathbf{E_7}$ \textbf{lattice}:
\begin{equation}
 E_7 = \left\{ v \in E_8 \, | \, v\cdot (0,0,0,0,0,0,1,-1) = 0\right\}\,. 
\end{equation}
The dual lattice $E_7^*$ contains 
\begin{equation}
v[1] = \left(\tfrac14,\tfrac14,\tfrac14,\tfrac14,\tfrac14,\tfrac14,-\tfrac34,-\tfrac34 \right)\,,
\end{equation}
and $E_7^*/E_7 = \Z_2$. The generator of the quotient obeys $v[1]\cdot v[1] = 3/2$. \\

$\mathbf{E_6}$ \textbf{lattice}:
\begin{equation}
 E_6 = \left\{ v \in E_8 \, | \, v\cdot (1,0,0,0,0,0,0,1) = v\cdot (\tfrac12,\tfrac12,\tfrac12,\tfrac12,\tfrac12,\tfrac12,\tfrac12,\tfrac12) =  0\right\} \,.
\end{equation}
The dual lattice $E_6^*$ contains the non-zero elements 
\begin{equation}
\begin{aligned}
v[1] &= \left(0,-\tfrac23,-\tfrac23,\tfrac13,\tfrac13,\tfrac13,\tfrac13,0 \right)\,, \\
v[2] &= \left(0,\tfrac23,\tfrac23,-\tfrac13,-\tfrac13,-\tfrac13,-\tfrac13,0 \right)  = -v[1]\,,
\end{aligned}
\end{equation}
and $E_6^*/E_6 = \Z_3$. The generator of the quotient obeys $v[1]\cdot v[1] = 4/3$. \\

For every root lattice, we can choose a $\Z$ basis composed of roots, such that the inner form $\Omega_{ij}$ between them can be read off from the Dynkin diagram of the associated Lie algebra: every node corresponds to a basis element, and two basis elements have inner form $-1$ if the two associated nodes are joined by a line. For the ADE lattices described above these diagrams are depicted as follows:

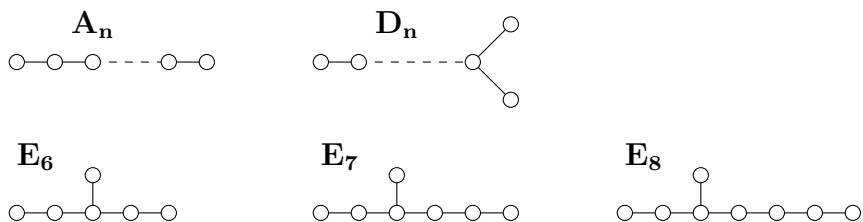
\hspace{0.75in}
\begin{tikzpicture}
	\draw(1,0.5)node{$\mathbf{A_n}$};
	\draw (0,0)--(1,0);
	\draw[dashed] (1,0)--(2,0);
	\draw (2,0)--(2.5,0);
	\draw[fill=white] (0,0) circle(0.1);
	\draw[fill=white] (0.5,0) circle(0.1);
	\draw[fill=white] (1,0) circle(0.1);
	\draw[fill=white] (2.0,0) circle(0.1);
	\draw[fill=white] (2.5,0) circle(0.1);
	
	\begin{scope}[shift={(4,0)}]
	\draw(1,0.5)node{$\mathbf{D_n}$};
	\draw (0,0)--(0.5,0);
	\draw[dashed] (0.5,0)--(2,0);
	\draw (2,0)--(2,0);
	\draw (2,0)--(2.5,0.5);	\centering
	\draw (2,0)--(2.5,-0.5);
	\draw[fill=white] (0,0) circle(0.1);
	\draw[fill=white] (0.5,0) circle(0.1);
	\draw[fill=white] (2.0,0) circle(0.1);
	\draw[fill=white] (2.5,0.5) circle(0.1);
	\draw[fill=white] (2.5,-0.5) circle(0.1);
	\end{scope}
	
	\begin{scope}[shift={(0,-2)}]
	\draw(0.25,0.75)node{$\mathbf{E_6}$};
	\draw (0,0)--(2,0);
	\draw (1,0)--(1,0.5);

	\draw[fill=white] (0,0) circle(0.1);
	\draw[fill=white] (0.5,0) circle(0.1);
	\draw[fill=white] (1,0) circle(0.1);
	\draw[fill=white] (1.5,0) circle(0.1);
	\draw[fill=white] (2.0,0) circle(0.1);
	\draw[fill=white] (1.0,0.5) circle(0.1);
	\end{scope}
	
	\begin{scope}[shift={(4,-2)}]
	\draw(0.25,0.75)node{$\mathbf{E_7}$};
	\draw (0,0)--(2.5,0);
	\draw (1,0)--(1,0.5);
	
	\draw[fill=white] (0,0) circle(0.1);
	\draw[fill=white] (0.5,0) circle(0.1);
	\draw[fill=white] (1,0) circle(0.1);
	\draw[fill=white] (1.5,0) circle(0.1);
	\draw[fill=white] (2.0,0) circle(0.1);
	\draw[fill=white] (2.5,0) circle(0.1);
	\draw[fill=white] (1.0,0.5) circle(0.1);
	\end{scope}
	
	\begin{scope}[shift={(8,-2)}]
	\draw(0.25,0.75)node{$\mathbf{E_8}$};
	\draw (0,0)--(3,0);
	\draw (1,0)--(1,0.5);
	
	\draw[fill=white] (0,0) circle(0.1);
	\draw[fill=white] (0.5,0) circle(0.1);
	\draw[fill=white] (1,0) circle(0.1);
	\draw[fill=white] (1.5,0) circle(0.1);
	\draw[fill=white] (2.0,0) circle(0.1);
	\draw[fill=white] (2.5,0) circle(0.1);
	\draw[fill=white] (3,0) circle(0.1);
	\draw[fill=white] (1.0,0.5) circle(0.1);
	\end{scope}
\end{tikzpicture}

\noindent For a specific choice of a $\Z$ basis composed of roots, the roots contained in this basis are called \emph{simple roots}. 

For any pair of root lattices, the existence of a primitive embedding of one into the other can be inferred from their Dynkin diagrams alone \cite{alma9944025973902959,nishiyama_1}: 
\begin{THM}
Up to automorphism, every primitive embedding between root lattices is given by an appropriate identification of simple roots.  
\end{THM}

To see this consider some primitively embedded root lattice $L$ into another root lattice $M$. Primitivity implies that $L$ is defined as the intersection of $M$ with some linear space of dimension equal to the rank of $L$, hence the simple roots of $L$ can be taken as simple roots of $M$.  For explicit classifications of such embeddings see \cite{nishiyama_1}.

\subsection{The Niemeier Lattices}\label{app:niemeier}

For rank $24$ there are $24$ even self-dual lattices of definite signature, which are called the \emph{Niemeier lattices} $N_I$. Following 
\cite{conway1998sphere}, we will denote them by the $24$ letters of the Greek alphabet. All of the $N_I$ except for the Leech lattice $\omega$ (which has no roots) can be constructed by 
starting with a direct sum of root lattices and adding in glue vectors. In table \ref{table:Niemeier}, we have collected the root sublattices of the $N_I$ and the glue vectors. For
\begin{equation}
(N_I)_\text{root} = \Gamma_1 \oplus \Gamma_2 \oplus \cdots   
\end{equation}
the glue vectors can be written as 
\begin{equation}
v_k = \left(v[a_k],v[b_k],\dots \right) 
\end{equation}
where $v[a_k] \in \Gamma_1^*, v[b_k] \in \Gamma_2^*, \cdots$ and so on. Using the notation introduced above, we will abbreviate this as $[a_1 b_1 \cdots]$.
We will use the notation $[(a_1 b_1 c_1)]$ to indicated that all cyclic permutations of glue vectors are used.

\begin{table}[h!]
\begin{center}
  \begin{tabular}{|c|c|c|}
  \hline
  $I$ & $(N_I)_\text{root}$ & glue vectors \\
  \hline
  \hline
   $\alpha$ & $D_{24}$& $[1]$\\
   $\beta$  & $D_{16}\oplus E_8$ &$[10]$ \\
   $\gamma$  & $E_8^3$ & --- \\
   $\delta$  & $A_{24}$ & $[5]$\\
   $\epsilon$  & $D_{12}^2$& $[(12)]$\\
   $\zeta$ & $A_{17} \oplus E_7$ & $[31]$\\
   $\eta$ & $D_{10}\oplus E_7^2$& $[110],[301]$\\
   $\theta$ &$A_{15} \oplus D_9$ & $[21]$\\
   $\iota$ & $D_8^3$& $[(122)]$\\
   $\kappa$ & $A_{12}^2$& $[15]$\\
   $\lambda$ & $A_{11}\oplus D_7 \oplus E_6$& $[111]$\\
   $\mu$ & $E_6^4$& $[1(012)]$\\
   $\nu$ & $A_9^2 \oplus D_6$& $[240],[501],[053]$\\
   $\xi$ & $D_6^4$& even permutations of $[0123]$\\
   $o$ & $A_8^3$& $[(114)]$\\
   $\pi$ & $A_7^2 \oplus D_5^2$ & $[1112],[1721]$\\
   $\rho$ & $A_6^4$& $[1(216)]$\\ 
   $\sigma$ & $A_5^4 \oplus D_4$& $[2(024)0],[33001],[30302],[30033]$\\
   $\tau$ & $D_4^6$& $[111111],[0(02332)]$\\
   $\upsilon$ & $A_4^6$ & $[1(01441)]$\\
   $\phi$ & $A_3^8$& $[3(2001011)]$\\
   $\chi$ & $A_2^{12}$& $[2(11211122212)]$\\
   $\psi$ & $A_1^{24}$& $[1(00000101001100110101111)]$\\
   $\omega$ & -- & --\\
   \hline
  \end{tabular}

\end{center}
\caption{The Niemeier lattices $N_I$.\label{table:Niemeier}}
\end{table}

\subsection{Embeddings into Niemeier Lattices}\label{app:embd_in_Niemeier}

Here, we will state some necessary conditions for embeddings of lattices $T_0$ of rank $6$ into a lattice $N$ of higher dimension. The application we have in mind is when $N$ is one of the Niemeier lattices. The simplifications we are interested in involve studying the root sublattice $(T_0)_\text{root}$ of $T_0$, which must be embedded into the root sublattice  $N_\text{root}$ of $N$.  

\begin{PR}\label{pr:OLADE6}
An even overlattice of a root lattice of rank $<8$ is again a root lattice. 
\end{PR}
{\it proof}: A root lattice $L$ has an even overlattice if there exists an element in its dual lattice $L^*$ not in $L$ and with even norm. It can be checked by inspection that for rank less than 8, all such elements have norm 2 and so are roots; e.g. $A_1^4 \subset D_4$. For rank 8 the latice $A_1^8$ has even overlattice given by the sum of fundamental weights of each $A_1$, which is a vector with norm 4. 

\begin{PR}
Let $\Lambda \subset M \subset N$ and assume that $\Lambda$ is not primitively embedded into $M$. Then $\Lambda$ is also not primitively embedded into $N$. 
\end{PR}
{\it proof}: As  $\Lambda$ is not primitively embedded into $M$ there is an $m \in M$ such that $m \notin \Lambda$ but $km \in \Lambda$ for $k \neq 1$. As $M$ is inside $N$ we again have $m \in N$, $m \notin \Lambda$ but $km \in \Lambda$, so that $\Lambda$ is also not primitively embedded in $N$. 

A consequence of this is that if $T_0$ is a root lattice primitively embedded into one of the Niemeier lattices $N_I$, it must necessarily be primitively embedded 
into $(N_I)_\text{root}$. However, not all of the lattice $T_0$ we are interested in are of this type. We can formulate a stronger statement for lattices of low rank however:

\begin{PR}\label{prop:prim}
Let $T$ be a lattice of rank $<8$ that is primitively embedded into a lattice $N$. Then $T_\text{root}$ is primitively embedded into $N_\text{root}$. 
\end{PR}
{\it proof}: To see this, let us assume that the embedding of $T$ into $N$ is primitive, but the embedding of $T_\text{root}$ into $N_\text{root}$ is not. 
Then there exists a $m \in N_\text{root}$ which is not in $T_\text{root}$ but $km \in T_\text{root}$ for $k \neq 1$. We can hence form an overlattice $T_\text{root}'$ 
of $T_\text{root}$ which contains $m$ as well. As we have seen in Proposition \ref{pr:OLADE6}, $T_\text{root}'$  must again be a root lattice. As $km \in T_\text{root}$
and $m \in N$, and furthermore $T$ is primitively embedded in $N$, it follows that $m \in T$ as well; otherwise 
$\left(T \otimes \Q\right) \cap N \neq T$. But this implies that $T_\text{root}' \subset T$, which is a contradiction as we assumed that $T_\text{root}$ 
is the root sublattice of $T$ and not $T_\text{root}$. 

As we are interested in lattice $T_0$ of rank $6$, every primitive embedding of $T_0$ into any one of the Niemeier lattice hence implies a primitive embedding 
of $(T_0)_\text{root}$ into $(N_I)_\text{root}$.

\begin{PR}\label{prop:roots}
Let $T_\text{root}$ be a root lattice and $N_I$ be a Niemeier lattice. Modulo automorphism, primitive embeddings of $T_\text{root}$ into $N_I$ are uniquely specified by 
an appropriate identification of the simple roots of $T_\text{root}$ and $(N_I)_\text{root}$. 
\end{PR}
{\it proof}: As stated above, primitive embeddings between root lattice are uniquely given by such an identification of simple roots, hence the same holds for 
primitive embeddings of $T_\text{root}$ into $(N_I)_\text{root}$. In other words for any primitive embedding we can use automorphisms of $(N_I)_\text{root}$ to achieve such a description. As the automorphism group of a Niemeier lattice $N_I$ contains the automorphism group of its root sublattice $(N_I)_\text{root}$ \cite{conway1998sphere}, 
we can use the same automorphism to bring any primitive embedding of $T_\text{root}$ into $N_I$ into such a form. 

\section{Examples of \texorpdfstring{$W_\text{root}$}{Wroot}}\label{app:alg}

In this appendix we record a list of possible root sublattices of frame lattices for $\K3$ surfaces with given transcendental lattice $T_S$ studied in this paper; see Tables \ref{tab:samp1} and \ref{tab:samp2}. We give up to 20 examples for each $T_S$, with at least one representative for each possible rank. The case $[6~1~1]$ is not included as it is exactly equivalent to $[3~1~2]$ (cf. Table \ref{tab:Dynkins}). These examples do not nearly exhaust every possibility and are recorded here for illustrative purposes. In Table \ref{tab:ranks} we give the full number of such lattices we have found in our exploration using the Kneser-Nishiyama method, organized into the allowed ranks. We do not claim exhaustivity of this exploration and so each number must be interpretted strictly as a lower bound.  

In the data below it is reflected that, as mentioned in the text, every simple ADE algebra of rank $\leq 18$ is allowed within the tadpole bound $\Nf \leq 24$. Note that we do not claim that these simple algebras appear in an isolated manner, i.e. as $\mathfrak{g}_r \oplus \mathfrak{u}_1^{18-r}$. The point is that at least at the level of ADE types, there is no forbidden Kodaira singularity in the elliptic fibrations under consideration.

\begin{table}[htbp]
	\footnotesize
	\begin{center}
		\begin{tabular}{||>{$}c<{$}|>{$}c<{$}||}
			\hline
			T_S & \text{Sample of algebras}\\ \hline
			\hline  1,0,1  &   A_{11} E_6,A_3 A_{15},D_6^3,D_6 D_{12},D_{18},D_4 E_7^2,A_1^2 E_8^2,A_9^2,A_1 A_{17},A_1^2 D_8^2,A_1^2 D_{16},A_1 D_{10} E_7,D_{10} E_8   \\
			\hline  1,1,1  &   A_{17},D_{10} E_7,A_{11} D_7,A_2 D_{16},E_6^3,A_2 E_8^2   \\
			\hline  2,0,1  &   A_7 A_9,A_9 D_7,A_5 A_7 D_5,A_1 D_8^2,A_1^2 A_9 E_6,A_1^2 D_8 E_7,A_1 A_3 A_7^2,A_1 D_7 D_{10},A_1 A_3 D_{14},A_1 A_3 E_7^2   \\
			&   A_1 A_{15},A_{17},A_1 D_4 D_6^2,A_1 D_6 D_{10},A_5 E_6^2,A_1 E_8^2,A_1 A_3 D_6 D_8,A_1 D_5 D_{12},A_1 D_{17},A_3 E_7 E_8   \\
			\hline  2,0,2  &   A_7^2,A_3 A_7 D_5,A_1^2 A_5^2 D_4,A_1^2 A_7 D_7,A_9 E_7,A_{11} D_6,D_{17},A_1 A_3 D_6 E_7,A_1^4 A_7^2,D_5 D_{13}   \\
			&   A_{15},A_{16},A_5^2 D_6,D_8^2,A_1 A_3 A_{13},A_1^2 A_3 D_6^2,A_7 D_4 E_6,A_3^6,A_1^2 A_3^2 D_{10},A_1 A_3 D_7 E_7   \\
			\hline  2,1,1  &   D_8^2,A_5 A_{12},A_{17},A_2 A_9 D_6,D_5 D_{12},D_{17},A_4 D_7 E_6,D_{10} E_7,D_9 E_8,A_6^3   \\
			&   A_1 A_8^2,A_1^2 A_{15},A_7 D_5^2,A_8 D_9,A_1 D_{16},A_{11} E_6,A_{10} E_7,A_3 E_7^2,A_1 E_8^2   \\
			\hline  2,1,2  &   A_1 A_7^2,A_7 D_4^2,A_1 D_7^2,A_3^2 D_5^2,A_4 D_5 D_7,A_5 D_{11},A_2^2 A_5 A_8,A_1 A_{11} D_5,D_{17},A_2^2 A_{14}   \\
			&   A_{15},A_{10} D_5,A_3 A_8 D_5,A_5 D_5 D_6,A_8 D_8,A_1 D_8 E_7,A_4^2 A_9,D_5 D_{12},A_2 A_8 E_7,A_2 A_9 E_7   \\
			\hline  2,2,2  &   A_{11} D_4,A_1^2 A_7^2,A_4 A_7 D_5,D_{16},E_8^2,A_1^3 A_{14},A_{12} D_5,D_4 D_5 D_8,A_{11} E_6,A_1^3 A_5 D_{10}   \\
			&   A_8 D_7,D_4^4,D_8^2,A_2^2 E_6^2,A_1^2 A_6 A_9,A_1^2 A_{15},A_1^2 A_3 D_6^2,D_5 D_{12},A_1^3 D_7 E_7,D_4 E_6 E_8   \\
			\hline  3,0,1  &   A_9 E_6,A_8^2,A_3 A_{13},A_{16},A_2 A_{15},A_3 A_5^2 D_4,A_1 D_8^2,A_1^2 A_{11} D_5,A_1^2 A_2 D_{14},A_1 A_2 D_8 E_7   \\
			&   A_2 A_6 A_8,A_7 A_9,A_1 A_{15},A_7 D_9,A_{17},A_1 D_4 D_6^2,A_1^2 D_8 E_7,A_2 D_6 D_{10},A_1 A_5 E_6^2,A_1 A_2 E_7 E_8   \\
			\hline  3,0,2  &   A_4^2 A_6,A_1 A_{13},A_1^2 A_4^2 A_5,A_5^2 D_5,A_1^2 A_2 A_6^2,A_1^3 D_7 E_6,A_1^2 D_7 D_8,A_1 A_9 E_7,A_3 A_5^3,A_1 A_2 D_5 D_{10}   \\
			&   A_7^2,A_4 A_6 D_4,A_{11} D_4,A_2 A_8 D_5,A_5 D_4 D_7,E_8^2,D_7 D_{10},A_1 A_3 D_6 E_7,A_1 A_2 A_{15},A_{11} E_7   \\
			\hline  3,0,3  &   A_2 A_3^2 A_5,A_3 A_5^2,A_3 A_5 A_6,A_3 A_4 A_7,A_1 A_5^2 D_4,A_1^2 A_3 D_4 D_6,D_7 D_9,A_1 A_{11} D_5,A_1^2 A_3 A_5 E_7,A_1^2 A_5^2 D_6   \\
			&   A_1 A_2 A_5^2,A_7 E_6,A_2 A_6^2,A_1 A_5 A_8,A_5 D_5^2,D_8^2,E_8^2,A_1 A_2 D_6 D_8,A_2^4 A_5^2,A_2 A_5^2 E_6   \\
			\hline  3,1,1  &   A_{15},A_6^2 D_4,A_5 D_5 D_6,A_2 D_7^2,A_1 D_9 E_6,A_1 A_4 A_5 A_7,A_1 A_{11} D_5,A_6 D_{11},A_1 D_{16},A_2 E_7 E_8   \\
			&   A_1 A_7 A_8,A_1 A_{10} D_5,A_9 D_7,D_6 D_{10},A_1 D_8 E_7,A_{17},A_2 A_7 D_8,A_1 A_2 D_{14},D_{10} E_7,A_{10} E_8   \\
			\hline  3,1,2  &   A_3 A_5 A_6,A_6 A_8,A_1 A_4 A_5^2,A_1 A_2 A_6 D_6,A_1 D_7 E_7,A_1^3 A_{13},A_2 D_5 D_9,A_1 A_2 E_6 E_7,D_{17},D_9 E_8   \\
			&   A_7^2,A_1 A_3^2 A_4^2,A_{15},A_1^2 A_7 E_6,A_1 A_3 A_6^2,A_1 A_{10} D_5,A_1 A_5 D_{10},A_{17},A_1 A_2 A_8 E_6,A_1 E_8^2   \\
			\hline  3,2,2  &   A_7^2,A_3 A_6^2,A_5^2 D_5,A_2 D_7 E_6,A_4^4,D_6 D_{10},A_1 A_{11} D_5,A_1 A_2 D_6 D_8,A_{11} E_6,A_4 E_7^2   \\
			&   A_5 A_9,A_1 A_5^2 D_4,A_1 A_7 D_7,A_2^2 A_4^3,A_1^2 A_7^2,A_1^2 E_7^2,A_1^2 A_3 D_6^2,A_1 A_2 D_{14},A_1^2 A_4 D_{12},A_4 D_6 E_8   \\
			\hline  3,3,3  &   A_3^3 A_5,A_{14},D_4^2 E_6,A_3 A_6^2,A_3^2 A_9,A_9 D_7,A_4 D_5 E_7,A_1 A_{11} D_5,A_2 A_4 D_{11},A_2^2 A_8 E_6   \\
			&   A_3 A_5 A_6,A_5^2 D_4,A_1 A_3^2 A_4^2,A_1^2 A_6 A_7,A_1 A_2^2 A_5 D_5,A_6 D_{10},E_8^2,A_2 A_7 D_8,A_1 A_2 D_{14},A_2 A_8 E_8   \\
			\hline  4,0,1  &   A_5^3,A_4 A_6 D_5,A_{10} D_5,A_9 E_6,A_1 A_7 A_8,A_1 A_7 D_4^2,A_3 E_6 E_7,A_1^2 A_3^2 A_5 D_4,A_1 D_4 D_{12},A_1^2 D_8 E_7   \\
			&   A_1 A_7^2,A_2 A_8 D_5,A_1 D_7^2,A_1 A_2 A_6 A_7,A_3 A_{13},A_4 A_6 E_6,A_1 A_4 A_{12},A_1^3 D_6 D_8,D_4 D_6 E_7,A_1 A_7 D_5^2   \\
			\hline  4,0,2  &   A_3 A_5^2,A_1 A_2 A_4 A_7,A_3 A_{11},A_4^2 A_7,A_1^4 A_7 D_4,A_8^2,A_1 A_5 D_4 D_6,A_1 A_3 A_{13},A_3 D_5 D_9,A_1^2 A_7 D_9   \\
			&   A_2 A_6 D_5,A_1 A_2^2 A_9,A_1^2 E_6^2,A_1 A_6 A_8,A_1^4 D_5 D_6,A_7 D_4 D_5,A_1^2 A_4^2 E_6,A_1^2 A_3 D_5 D_7,D_{17},A_1 A_3 A_7 E_7   \\
			\hline  4,0,4  &   A_2^4 A_3,A_1 A_5 D_5,A_1^2 A_3^2 D_4,A_1 A_2 A_{10},A_1^2 D_5 D_6,A_1 A_2 A_4 A_7,A_1 A_2 A_3 A_4 A_5,A_1 A_7^2,D_4 D_6^2,A_1^2 A_3 A_7 D_5   \\
			&   A_3 A_4^2,A_1^8 A_2^2,D_6^2,A_1^2 A_3^2 D_5,A_2^4 A_3^2,A_1^4 D_5^2,A_1 A_3 A_5 A_6,A_3 A_7 D_6,A_9 E_7,A_7 D_5^2   \\
			\hline  4,1,1  &   A_6 A_9,A_3 D_6^2,A_2 A_{14},A_2 A_4 D_5^2,A_9 E_7,A_1^2 E_7^2,A_7 D_5^2,A_4 D_{13},A_1 A_2^2 E_6^2,A_1 E_8^2   \\
			&   A_{15},A_7 A_9,A_{16},A_1 A_8 D_7,A_2 D_7 E_7,A_{17},A_2 A_5 D_{10},A_1 D_{16},D_{10} E_7,A_4 E_6 E_8   \\
			&   A_1^2 A_3 A_4 A_5,A_1 A_5 A_8,A_1 A_7 D_6,A_1 A_2 A_6^2,A_{11} D_4,A_1 A_4^2 A_7,A_4 A_{12},A_1 A_3 D_{12},A_1 D_{16},A_4 D_5 E_8   \\
			\hline  4,2,2  &   A_1 A_3 A_5^2,A_3 A_4 A_7,A_1^2 E_6^2,A_1^2 A_6 A_7,A_1 A_3 A_5 D_6,A_3 A_5 A_8,A_1^3 A_9 D_4,A_{10} E_6,A_1 D_6 D_{10},D_{10} E_7   \\
			&   A_1^2 A_6^2,A_1^2 A_5 A_7,A_2^3 A_3^3,A_1 A_2 A_3 A_9,A_3 D_5 D_7,A_1^2 A_2 A_3 A_5 D_4,A_3 A_4 D_9,A_1^3 D_6 D_8,A_{11} E_6,D_4 D_5 E_8   \\
			\hline  4,2,4  &   A_1 A_2 A_4^2,A_1^2 A_5 D_4,A_1^3 A_3^3,A_3^3 A_4,A_1^4 A_3 A_6,A_1^3 A_5 A_6,A_1^5 A_3 A_7,A_1^3 A_2^2 A_9,A_1^4 A_3^2 A_7,A_4 D_8 E_6   \\
			&   A_1 A_4 A_6,A_2^3 A_3^2,A_1^2 A_2 A_4^2,A_1^3 A_2 A_3 A_5,A_1^2 A_2 A_3^2 A_4,A_1 A_2 A_4 A_7,A_3^2 A_4 D_5,A_{10} E_6,A_1^2 A_2 A_5 D_8,A_1^2 A_4 A_5 E_7   \\
			\hline  4,4,4  &   A_2^6,A_2^2 D_4^2,A_1 A_2 A_3 A_7,A_1^2 A_3 A_4 D_4,A_4^2 A_6,A_3 A_5 E_6,A_3 D_5 D_7,A_1 A_7 D_4^2,A_1 A_9 D_6,A_1 A_3^2 D_5^2   \\
			&   A_4^3,E_6^2,A_3^3 D_4,A_3 A_5 D_5,A_1 A_3^2 A_7,A_1^2 A_5 D_4^2,A_1^2 A_3 A_4 E_6,A_1 A_3^3 D_6,A_1^4 A_3^2 A_7,A_7 D_5^2   \\
			\hline  5,0,1  &   A_8 D_6,A_6 A_9,A_3 D_6^2,A_4 D_5 E_6,A_1 A_{15},A_3 D_5 D_8,A_1^2 D_8 E_6,A_1 D_8^2,A_1 D_8 E_8,A_1 E_8^2   \\
			&   A_4 A_5 A_6,A_2^2 A_7 D_4,A_2 A_6 D_7,A_7 A_9,A_4 A_7 D_5,A_1^2 D_5 D_9,A_1^3 A_3 A_5 D_6,A_1 D_{16},A_1^2 E_7 E_8,A_1 A_9 D_8   \\
			\hline  5,0,5  &   A_1^2 A_2^2 A_3,A_1^4 A_2 A_4,A_1^3 A_2^2 A_4,A_2^6,A_1 A_3 A_4 A_5,A_1^2 A_2 A_3 A_7,A_1 A_3 D_{11},A_1^3 A_6 D_7,A_1 A_9 E_7,A_1^2 A_4^4   \\
			&   A_2^2 A_3^2,A_1^2 A_4^2,A_{11},A_1^8 D_4,A_3^3 D_4,A_1 A_4 D_4 D_5,A_4 D_{11},A_1^4 D_{12},A_1 A_4 D_5 E_7,A_1^2 A_4^2 E_8   \\\hline
		\end{tabular}		
	\end{center}
\caption{Some examples of root systems of gauge algebras appearing for $\K3$ surfaces with transcendental lattice $T_S$.}
\label{tab:samp1}
\end{table}

\begin{table}[htbp]
	\footnotesize
	\begin{center}
		\begin{tabular}{||@{}>{$}c<{$}@{}|>{$}c<{$}||}
			\hline
			T_S & \text{Sample of algebras}\\ \hline
			\hline  5,1,1  &   A_5^3,A_1^2 A_8 D_5,A_{10} D_5,A_1 D_7^2,A_2^2 A_4^2 D_4,A_5^2 E_6,A_2 E_7^2,A_3 A_5 A_9,A_3 A_{14},A_2 D_7 E_8   \\
			&   A_4^2 A_7,A_1 A_9 D_5,A_1 A_5 D_4 D_5,A_1^2 A_4 A_{10},A_1 A_4 A_6 D_5,A_2 A_8 E_6,A_4^2 A_9,A_3^2 A_{11},D_{10} E_7,A_{18}   \\
			\hline  5,5,5  &   A_2^5,A_1^5 A_3^2,A_1 A_2 A_4^2,A_2^3 A_3^2,A_1^4 A_2^3 A_3,A_1^2 A_2 A_4 D_5,A_2^3 D_4^2,A_1 A_2 A_3^2 D_6,E_8^2,A_4 A_5 E_8   \\
			&   A_2^3 D_4,A_2 A_3^3,A_1^2 A_2^2 A_3^2,A_1^2 A_2^3 A_4,A_1^3 A_3 A_7,A_1^2 A_4 A_8,A_1 A_2^4 D_6,A_1 A_2^2 A_5 E_6,A_4 E_6 E_7,A_4 A_5 A_9   \\
			\hline  6,0,1  &   A_1^2 A_5 A_7,A_2 A_7 D_5,A_1 A_5^2 D_4,A_2 A_7 D_6,A_3 A_4 A_9,A_1^2 A_4 A_5 D_5,A_1 D_{15},A_1 D_8^2,A_1^2 E_7 E_8,D_5 E_6 E_7   \\
			&   A_4 A_5 D_5,A_1 A_2 A_3 A_5 D_4,A_4 A_5 D_6,A_1^3 D_6^2,A_1 A_{15},A_1 A_3 D_{12},A_{17},A_1 D_7 D_9,A_1^2 A_3 A_5 D_8,A_5 D_5 E_8   \\
			\hline  6,0,2  &   A_3^4,A_6^2,A_1^2 A_4 A_7,A_1^2 A_6^2,A_1^2 D_6^2,A_3 A_8 D_4,A_1^6 A_3 A_7,A_1^2 A_5 A_9,A_1^2 A_3 D_{12},A_1 A_3 A_5 D_9   \\
			&   A_1^2 A_5^2,A_3^2 A_7,A_3^3 D_4,A_1 A_6 A_7,A_1^3 A_4^3,A_2^2 A_5 E_6,A_1^2 A_6 A_8,A_1^2 D_5 D_{10},A_3 D_6 E_8,D_5 D_7 E_6   \\
			\hline  6,0,3  &   A_2 A_4^2,A_1^3 A_2 A_3^2,A_1^4 A_2^4,A_2^3 A_3^2,A_1^2 A_2^2 A_7,A_2^2 A_4 A_6,A_1 A_2 A_3 A_4 D_5,A_1^2 A_2^4 A_6,D_5 D_6 E_6,A_2^2 A_3 A_5 E_6   \\
			&   A_1^5 A_3^2,A_1^3 D_4^2,A_1 A_2^4 A_3,A_1^3 A_2 A_4^2,A_1 A_2 A_3 A_7,A_8 D_6,A_1^3 A_3^2 D_6,A_5 A_{11},A_2^2 E_6 E_7,A_2 A_3 E_6 E_7   \\
			\hline  6,0,6  &   A_2^4,A_1 A_2 A_6,A_1^5 A_5,A_1 A_4 A_6,A_1^2 A_4 E_6,A_2 A_6 D_5,A_1^3 A_3^2 A_5,A_1 A_2 A_3 A_9,A_1^2 A_2^2 A_3^2 A_4,A_2 A_3 E_6^2   \\
			&   A_4^2,A_1 A_4 D_4,A_1^2 A_4 D_4,A_5 E_6,A_1^2 D_4 E_6,A_1 A_7 D_5,A_4 A_6 D_4,D_4 D_5 D_6,A_2 D_7 E_7,A_3^2 E_6^2   \\
			\hline  6,2,2  &   A_3^4,A_2 A_3 A_4^2,A_2 A_3 A_4 D_4,A_1 A_4^2 A_5,A_1 A_2 A_5 D_6,A_3 A_4^3,A_2 A_3 D_5^2,A_1^2 A_3 A_{11},A_3 D_{13},A_1^2 A_3 A_5 E_7   \\
			&   A_1^2 A_5^2,A_2 A_3 A_8,A_1^2 A_2 A_5 D_4,A_1^6 D_4^2,A_1^2 D_6^2,A_1^2 A_3 A_5 D_5,A_1 A_7 D_7,A_1^2 A_9 D_5,A_1^2 A_{15},D_9 E_8   \\
			\hline  6,3,3  &   A_1^2 A_3^3,A_1^3 D_4^2,A_1 A_2 A_3^3,A_1^2 A_2^4 A_3,A_1 A_{12},A_1 D_6 D_7,A_3 A_7 D_5,A_2^8,A_2^2 E_6 E_7,A_1 A_2^2 A_{13}   \\
			&   A_2^2 A_7,A_2^3 A_3^2,A_1 A_2 A_4 D_5,A_1^3 A_2 A_8,A_1 A_3 A_6 D_4,A_2 A_4 D_8,A_8 D_7,A_1 A_5 A_{10},A_2 E_7 E_8,A_2^2 A_6 E_8   \\
			\hline  6,6,6  &   A_2^4,A_1^3 A_2^3,A_1 A_2^3 A_3,A_1^2 A_3^3,A_1^5 A_2 A_5,A_1^9 D_4,A_1^3 A_2 A_4 A_5,A_1 A_3 A_4 A_7,A_{12} D_4,A_1^2 A_2 A_3 A_5^2   \\
			&   D_4^2,A_1 A_4^2,A_1^2 A_2^2 A_4,A_1^2 A_4 D_5,A_1 D_5 D_6,A_2 A_5 D_6,A_1^2 A_3 A_4 A_5,A_1 A_7 D_7,A_1 A_5^2 D_5,A_2 A_5^2 D_5   \\
			\hline  7,0,1  &   A_1 A_2 A_5 A_6,A_4 A_6 D_4,A_1 A_4^2 D_5,A_5^3,A_5^2 D_5,A_1 A_2 A_5 A_8,A_1^3 A_9 D_4,A_2 A_7 E_7,A_1^2 D_7 D_8,A_1 D_{16}   \\
			&   A_7^2,A_1 A_5 D_4^2,A_1^2 A_4^2 A_5,A_1^3 A_3 A_5 D_4,A_1 A_5 D_4 D_5,A_5 A_{11},A_1^2 D_4 D_{10},A_1 A_4 A_6^2,A_1^2 D_5 D_{10},A_1 D_8 E_8   \\
			\hline  7,0,7  &   A_1^6,A_1^2 A_2 A_3,A_1^6 A_2,A_1^7 A_2,A_1^{10},A_1^2 A_3^3,A_1^4 A_2^4,A_2^2 A_3^3,A_1^2 A_2^2 A_4^2,A_3^5   \\
			&   A_1^2 A_2^2,A_2^2 A_3,A_1^4 A_2^2,A_1^5 A_4,A_1^7 A_3,A_1 A_3^2 A_4,A_1^2 A_2 A_4^2,A_2^3 A_3 A_4,A_1 A_2 A_3 A_4^2,A_1^3 D_{13}   \\
			\hline  7,1,1  &   A_2 A_3 A_4 A_5,A_1 A_8 D_5,A_8 E_6,A_1 A_4 A_5 D_5,A_6 D_9,A_2 A_3 A_5 A_6,A_2 A_4 A_5 D_5,A_1 D_7 E_8,A_2 A_5 D_{10},A_1 A_8 E_8   \\
			&   A_4 A_5^2,D_4 D_5^2,A_4 A_5 A_6,A_1^2 A_7 D_6,A_3 A_6 E_6,A_1 A_2 A_9 D_4,A_1 A_2 A_5 D_8,A_3 A_7 D_7,A_1 A_2^2 E_6^2,A_1 E_8^2   \\
			\hline  7,7,7  &   A_2^3,A_1^2 A_2^3,A_1^2 A_2 A_5,A_1 A_3^3,A_1^5 A_2 D_4,A_1 A_2^2 A_3 D_4,A_1^5 A_4^2,D_7 E_7,A_1^2 A_2 D_{11},A_2^2 D_{12}   \\
			&   A_1^6 A_2,A_1^4 A_5,A_1^5 A_2 A_3,A_1^5 A_2 A_4,A_2^3 A_3^2,A_2^5 A_3,A_1^4 A_5^2,A_3^3 A_6,A_4 A_6^2,A_1^2 A_2 A_{13}   \\
			&   A_1^2 A_3 A_4 A_5,A_1 A_3 A_{10},D_7^2,A_2 A_4 D_4 D_5,A_4 D_4 D_7,A_2^2 A_3 A_9,A_1 A_2 A_3 D_{10},A_1 D_8 E_7,A_1 A_2 A_{14},A_4 D_5 E_8   \\
			\hline  8,8,8  &   A_2^2,A_2^3,A_1^5 A_2,A_1^3 A_2 A_3,A_1^3 A_2 D_4,A_2^2 A_3^2,A_2 A_3^3,A_1 A_2 A_3 A_6,A_1 A_2 A_5^2,A_1^2 A_3^3 D_4   \\
			&   A_1^4 A_2,A_1^7,A_2^4,A_1 A_4^2,A_1 A_2^3 A_3,A_1^9 A_2,A_1 A_2^2 A_3 A_4,A_1 A_2^3 A_3^2,A_2^3 A_4^2,A_1 A_{15}   \\
			\hline  9,9,9  &   A_1^3 A_2,A_1^3 A_3,A_1^5 A_2,A_1^8,A_1^3 A_3^2,A_1^7 A_3,A_1^4 A_2^2 A_3,A_2^2 A_3 A_5,A_2^3 A_3 A_4,A_1^3 A_2 A_5^2   \\
			&   A_1 A_2^2,A_2 D_4,A_1 A_2^3,A_1^2 A_2^3,A_2 A_3 D_4,A_3^2 D_4,A_1 A_2 A_3 A_5,A_1^6 A_2 D_4,A_4^2 A_6,A_2^8   \\
			\hline  10,10,10  &   A_1^4,A_1^3 A_2,A_1 A_2 A_3,A_1^2 A_2 A_3,A_1 A_2 A_5,A_1^3 A_6,A_1 A_2^2 E_6,A_1^5 A_2 A_6,A_1^2 A_4^2 A_5,A_1^3 A_2 A_4 E_8   \\
			&   A_1^5,A_1^6,A_1^4 A_3,A_4^2,A_2^3 A_3,A_1^6 A_2^2,A_6^2,A_1 A_5 E_8,A_1 A_2 A_4 A_9,A_1^3 A_2 A_4 A_9   \\ \hline
		\end{tabular}		
		\caption{Continuation of Table \ref{tab:samp1}.}
		\label{tab:samp2}
	\end{center}
\end{table}

\begin{table}[htbp]
	\footnotesize
	\begin{center}
		\begin{tabular}{||@{}>{$}c<{$}@{}|>{$}c<{$}|>{$}c<{$}|>{$}c<{$}|>{$}c<{$}|>{$}c<{$}|>{$}c<{$}|>{$}c<{$}|>{$}c<{$}|>{$}c<{$}|>{$}c<{$}|>{$}c<{$}|>{$}c<{$}|>{$}c<{$}|>{$}c<{$}|>{$}c<{$}|>{$}c<{$}|>{$}c<{$}|>{$}c<{$}|>{$}c<{$}|>{$}c<{$}||}
			\hline
			T_S & 1 & 2 & 3 & 4 & 5 & 6 & 7 & 8 & 9 & 10 & 11 & 12 & 13 & 14 & 15 & 16 & 17 & 18 \\ \hline
			1,0,1  & 0 & 0 & 0 & 0 & 0 & 0 & 0 & 0 & 0 & 0 & 0 & 0 & 0 & 0 & 0 & 0 & 1 & 12 \\
			1,1,1  & 0 & 0 & 0 & 0 & 0 & 0 & 0 & 0 & 0 & 0 & 0 & 0 & 0 & 0 & 0 & 0 & 2 & 4 \\
			2,0,1  & 0 & 0 & 0 & 0 & 0 & 0 & 0 & 0 & 0 & 0 & 0 & 0 & 0 & 0 & 0 & 3 & 13 & 11 \\
			2,0,2  & 0 & 0 & 0 & 0 & 0 & 0 & 0 & 0 & 0 & 0 & 0 & 0 & 0 & 1 & 2 & 30 & 17 & 9 \\
			2,1,1  & 0 & 0 & 0 & 0 & 0 & 0 & 0 & 0 & 0 & 0 & 0 & 0 & 0 & 0 & 0 & 1 & 17 & 1 \\
			2,1,2  & 0 & 0 & 0 & 0 & 0 & 0 & 0 & 0 & 0 & 0 & 0 & 0 & 0 & 0 & 5 & 24 & 20 & 2 \\
			2,2,2  & 0 & 0 & 0 & 0 & 0 & 0 & 0 & 0 & 0 & 0 & 0 & 0 & 0 & 0 & 2 & 7 & 16 & 2 \\
			3,0,1  & 0 & 0 & 0 & 0 & 0 & 0 & 0 & 0 & 0 & 0 & 0 & 0 & 0 & 0 & 1 & 15 & 16 & 6 \\
			3,0,2  & 0 & 0 & 0 & 0 & 0 & 0 & 0 & 0 & 0 & 0 & 0 & 0 & 0 & 10 & 33 & 61 & 52 & 12 \\
			3,0,3  & 0 & 0 & 0 & 0 & 0 & 0 & 0 & 0 & 0 & 0 & 0 & 0 & 4 & 41 & 15 & 30 & 18 & 9 \\
			3,1,1  & 0 & 0 & 0 & 0 & 0 & 0 & 0 & 0 & 0 & 0 & 0 & 0 & 0 & 0 & 1 & 9 & 18 & 1 \\
			3,1,2  & 0 & 0 & 0 & 0 & 0 & 0 & 0 & 0 & 0 & 0 & 0 & 0 & 0 & 3 & 50 & 65 & 19 & 0 \\
			3,2,2  & 0 & 0 & 0 & 0 & 0 & 0 & 0 & 0 & 0 & 0 & 0 & 0 & 0 & 2 & 25 & 20 & 22 & 3 \\
			3,3,3  & 0 & 0 & 0 & 0 & 0 & 0 & 0 & 0 & 0 & 0 & 0 & 0 & 0 & 10 & 39 & 20 & 12 & 2 \\
			4,0,1  & 0 & 0 & 0 & 0 & 0 & 0 & 0 & 0 & 0 & 0 & 0 & 0 & 0 & 0 & 7 & 30 & 36 & 1 \\
			4,0,2  & 0 & 0 & 0 & 0 & 0 & 0 & 0 & 0 & 0 & 0 & 0 & 0 & 2 & 30 & 48 & 53 & 19 & 2 \\
			4,0,4  & 0 & 0 & 0 & 0 & 0 & 0 & 0 & 0 & 0 & 0 & 3 & 39 & 74 & 124 & 103 & 69 & 3 & 0 \\
			4,1,1  & 0 & 0 & 0 & 0 & 0 & 0 & 0 & 0 & 0 & 0 & 0 & 0 & 0 & 0 & 3 & 28 & 9 & 1 \\
			4,2,2  & 0 & 0 & 0 & 0 & 0 & 0 & 0 & 0 & 0 & 0 & 0 & 0 & 0 & 19 & 47 & 65 & 35 & 0 \\
			4,2,4  & 0 & 0 & 0 & 0 & 0 & 0 & 0 & 0 & 0 & 0 & 3 & 30 & 92 & 138 & 158 & 101 & 37 & 2 \\
			4,4,4  & 0 & 0 & 0 & 0 & 0 & 0 & 0 & 0 & 0 & 0 & 0 & 5 & 18 & 48 & 51 & 41 & 4 & 0 \\
			5,0,1  & 0 & 0 & 0 & 0 & 0 & 0 & 0 & 0 & 0 & 0 & 0 & 0 & 0 & 1 & 16 & 23 & 23 & 1 \\
			5,0,5  & 0 & 0 & 0 & 0 & 0 & 0 & 0 & 0 & 1 & 22 & 58 & 115 & 163 & 260 & 219 & 128 & 17 & 3 \\
			5,1,1  & 0 & 0 & 0 & 0 & 0 & 0 & 0 & 0 & 0 & 0 & 0 & 0 & 0 & 0 & 22 & 63 & 23 & 1 \\
			5,5,5  & 0 & 0 & 0 & 0 & 0 & 0 & 0 & 0 & 0 & 2 & 13 & 55 & 107 & 159 & 113 & 50 & 9 & 1 \\
			6,0,1  & 0 & 0 & 0 & 0 & 0 & 0 & 0 & 0 & 0 & 0 & 0 & 0 & 0 & 3 & 35 & 26 & 20 & 3 \\
			6,0,2  & 0 & 0 & 0 & 0 & 0 & 0 & 0 & 0 & 0 & 0 & 0 & 5 & 43 & 92 & 122 & 105 & 49 & 3 \\
			6,0,3  & 0 & 0 & 0 & 0 & 0 & 0 & 0 & 0 & 0 & 1 & 11 & 50 & 120 & 138 & 99 & 68 & 27 & 3 \\
			6,0,6  & 0 & 0 & 0 & 0 & 0 & 0 & 0 & 8 & 20 & 48 & 82 & 141 & 182 & 237 & 209 & 145 & 33 & 9 \\
			6,1,1  & 0 & 0 & 0 & 0 & 0 & 0 & 0 & 0 & 0 & 0 & 0 & 0 & 0 & 3 & 50 & 65 & 19 & 0 \\
			6,2,2  & 0 & 0 & 0 & 0 & 0 & 0 & 0 & 0 & 0 & 0 & 0 & 2 & 28 & 77 & 53 & 44 & 19 & 0 \\
			6,3,3  & 0 & 0 & 0 & 0 & 0 & 0 & 0 & 0 & 0 & 0 & 5 & 25 & 92 & 143 & 85 & 43 & 20 & 3 \\
			6,6,6  & 0 & 0 & 0 & 0 & 0 & 0 & 0 & 2 & 5 & 19 & 52 & 109 & 150 & 178 & 150 & 88 & 24 & 0 \\
			7,0,1  & 0 & 0 & 0 & 0 & 0 & 0 & 0 & 0 & 0 & 0 & 0 & 0 & 0 & 21 & 67 & 66 & 35 & 0 \\
			7,0,7  & 0 & 0 & 0 & 0 & 0 & 2 & 8 & 15 & 18 & 21 & 24 & 25 & 19 & 10 & 2 & 6 & 0 & 0 \\
			7,1,1  & 0 & 0 & 0 & 0 & 0 & 0 & 0 & 0 & 0 & 0 & 0 & 0 & 0 & 13 & 80 & 65 & 12 & 0 \\
			7,7,7  & 0 & 0 & 0 & 0 & 0 & 1 & 0 & 11 & 25 & 38 & 53 & 66 & 70 & 65 & 26 & 13 & 1 & 0 \\
			8,8,8  & 0 & 0 & 0 & 1 & 0 & 3 & 6 & 18 & 28 & 40 & 56 & 71 & 77 & 57 & 20 & 7 & 0 & 0 \\
			9,9,9  & 0 & 0 & 0 & 0 & 2 & 7 & 7 & 17 & 28 & 40 & 46 & 60 & 58 & 29 & 3 & 1 & 0 & 0 \\
			10,10,10  & 0 & 0 & 0 & 1 & 3 & 7 & 11 & 26 & 40 & 72 & 112 & 167 & 224 & 282 & 201 & 95 & 12 & 1 \\ \hline
		\end{tabular}		
		\caption{Numbers of root lattices found for a given $T_S$ with a certain rank. We have not aimed at exhaustivity and every entry must be interpreted strictly as a lower bound.}
		\label{tab:ranks} 
	\end{center}
\end{table}

\newpage
\bibliographystyle{utphys}
\bibliography{refs}

\end{document}